# Electrostatic Screening Modulation of Graphene's Electronic Structure and the Helical Wavefunction-Dominated Topological Properties

Yaorui Tan ,Xiang Chen, Yunhu Zhu, Xiaowu Yang, Zhongkai Huang, Chuang Yao, Maolin Bo*

*Key Laboratory of Extraordinary Bond Engineering and Advanced Materials Technology (EBEAM) of Chongqing, College of Mechanical and Electrical Engineering, Yangtze Normal University, Chongqing 408100, China*

*Corresponding author: Maolin Bo (E-mail addresses: bmlwd@yznu.edu.cn)

## Abstract

This study examines electrostatic screening effects in graphene using tight-binding calculations based on the Binding-energy and Bond-Charge (BBC) model and a modified version of it. The results indicate that the modified BBC potential decays in an exponential manner with distance, which suppresses electron–electron interactions. The hopping integrals exhibit a pronounced decrease over distance and shift with parameter variation, while the on-site energy increases linearly. A band gap opens once the parameter exceeds a certain threshold. The density of states shows a prominent peak near the Fermi level, whereas the low-energy region remains largely unchanged. The low-energy helical wave functions in graphene display topological characteristics, including pseudospin–momentum locking and a π Berry phase, resulting in distinctive transport properties. By avoiding the Coulomb singularity, the model offers valuable insights for the engineering of screening in two-dimensional systems and the design of topological devices.

## 1. Introduction

Since its discovery, graphene has become a major focus in condensed matter physics and materials science due to its unique two-dimensional hexagonal honeycomb lattice structure [1-7]. Its electronic bands exhibit a linear dispersion relation near the Dirac points, giving low-energy quasiparticles the characteristics of massless Dirac fermions and a zero-bandgap feature at the Fermi level . These intrinsic properties grant graphene ultra-high carrier mobility up to $2\times10^5$ $cm^2\cdot V^{-1}\cdot s^{-1}$ at room temperature, excellent electrical transport performance, and unique topological attributes. Consequently, it holds significant potential for applications in nanoelectronic devices, quantum information processing, and photodetection[8].

However, the zero-bandgap characteristic of graphene limits its application in core devices like logic switches and digital circuits, creating a critical bottleneck for industrial adoption[9]. As a prototypical two-dimensional material, graphene's electronic structure is susceptible to the external charge environment. The electrostatic screening effect from electron-electron interactions directly modulates its band dispersion, topological properties, and transport behavior[10]. Therefore, precise modulation of graphene's electronic structure through control of charge screening strength—particularly the tunable opening of a bandgap and flexible adjustment of topological properties—has become a central research focus and cutting-edge direction in this field[11].

The tight-binding (TB) model is widely used to describe the electronic structure of lattice materials such as graphene, owing to its clear physical interpretation and high computational efficiency[12-15]. Nevertheless, the conventional TB model has significant limitations. It often simplifies by considering only nearest-neighbor hopping integrals, failing to capture the non-local screening effects of electron-electron interactions and typically using the bare Coulomb potential, leading to short-range singularity issues. It also neglects the influence of overlap integrals and band renormalization effects on electronic responses and lacks a systematic exploration of the correlation between screening modulation and topological properties. Although some studies have attempted to introduce screening potential corrections[16], existing models still require improvement in physical self-consistency, quantitative accuracy of parameter modulation, and the coupled description of topological properties.

In addition, the charge screening effect is generally not explicitly considered in traditional density functional theory(DFT) calculations. In practical material calculations, charge screening is a key physical parameter characterizing the strength of electronic interactions, which directly determines the electronic polarization behavior of the system. Moreover, the charge screening effect varies significantly among different energy levels and orbitals, and thus cannot be simply neglected.Accordingly, this work introduces and systematically incorporates the charge screening effect into the calculation, improving the original theoretical framework. Although the $V_{BBC}$ model adopted in the present study shows certain differences in accuracy compared with conventional DFT calculations, future efforts can combine machine learning approaches to further compensate for the shortcomings of the current model in describing electronic interactions, thereby improving its accuracy and applicability.

Conventional tight-binding models usually exhibit nonlinear dispersion near the Fermi level, with physical mechanisms buried in numerical results and topological properties requiring extra computation, thus lacking intuitive analytical insight. In contrast, the helical wave model offers high computational efficiency and a transparent physical picture, providing analytical solutions near the Dirac point that directly yield topological quantities such as Berry curvature and helical phase while explicitly manifesting valley degrees of freedom and chiral symmetry, which greatly simplifies low-energy physics analysis. However, its validity is strictly restricted to the small-momentum regime (approximately $|q| < 0.2 \times 2\pi/a$), as it neglects band nonlinearity and electron interactions and cannot describe high-energy excitations, defect effects, or complex material modifications.

To address the aforementioned issues, this study introduces the Binding-energy and Bond-Charge(BBC) and improved BBC screening potential functions within the tight-binding model framework[17]. We construct a self-consistent screened tight-binding model to investigate the regulatory mechanisms of electrostatic screening effects on graphene's electronic structure and topological properties. The core content includes calculating key physical quantities—such as band structure, electronic density of states (DOS), helical wavefunctions[18], and Berry curvature[19-21]—for graphene under varying screening parameters $\sigma_v$; analyzing the evolution of screening potential, hopping integrals, and on-site energy with interatomic distance and $\sigma_v$; and revealing the intrinsic correlation between screening effects and topological properties.

This study provides a systematic theoretical framework for understanding how electrostatic screening modulates electron-electron interactions, band renormalization, and topological properties in graphene. It also offers crucial theoretical support for the tunability of Fermi velocity, bandgap size, and carrier type through experimental regulation of $\sigma_v$ using methods such as electrostatic gating, substrate engineering, or chemical doping. The findings are significant for optimizing performance and innovating designs of graphene-based high-speed transistors, tunable photodetectors, and topological quantum devices. They also serve as quantitative references for advancing screening engineering in two-dimensional materials and developing distance-modulated quantum devices.

## 2. Principles and Calculation Methods

### 2.1 Tight-Binding approximation and BBC potential

Using the tight-binding approximation to describe the electronic structure of graphene, the spatial Hamiltonian can be expressed as:

$$\hat{H}=\sum_{m,R}\varepsilon_m \hat{c}^{\dagger}_{m,R}\hat{c}_{m,R}+\sum_{m,n,R,R'}t_{mn}(R-R')\hat{c}^{\dagger}_{m,R}c_{n,R'}+h.c. \tag{1}$$

Among them, $\hat{c}^{\dagger}_{m,R}$ and $\hat{c}_{m,R}$ respectively represent the electron generation and annihilation operators on orbital $\alpha$ at lattice position $i$; $\varepsilon_m$ is the energy of atomic orbitals; $t_{mn}(R-R')$ is the jump integral, which describes the coupling strength between different atomic orbitals; Electron generation and annihilation operators on $R$ and $R'$ orbitals $m$ and $n$ ; $h.c.$ represents the Hermitian conjugate term.

The Bloch wave function is obtained by linearly combining atomic orbital wave functions:

$$\psi_{mk}(r)=\frac{1}{\sqrt{N}}\sum_{R}e^{ik\cdot R}\phi_m(r-R) \tag{2}$$

Among them, $N$ is the total number of cells, is the $m$-th atomic orbital wave function, and $k$ is the inverted lattice space wave vector.

The core of the BBC potential is the Coulomb interaction between shielded atoms, which takes the basic form of [22]:

$$V_{\text{BBC}}(\vec{R})=\frac{1}{4\pi\epsilon_0}\cdot\frac{\left(Z-\sigma_v\right)e^2}{\left|\vec{R}\right|} \tag{3}$$

Among them, $\frac{1}{4\pi\epsilon_0}$ is the Coulomb constant; $e^2$ is the Square of the electronic charge; $\left|\vec{R}\right|$ is the interatomic distance ($\mathring{A}$).

The BBC model describes the shielding effect of electron-electron interactions on the Coulomb potential by introducing shielding parameters. The energy of atomic orbitals is determined by both the kinetic energy term and the BBC shielding

potential:

$$\varepsilon_m = \langle \phi_m \,|\, \hat{T} + \hat{V}_{\mathrm{BBC}} \,|\, \phi_m \rangle \tag{4}$$

Where $\hat{T} = -\frac{h^2}{2m}\nabla^2$ is the kinetic energy operator and $\hat{V}_{\mathrm{BBC}}$ is the BBC shielding potential.

The specific form of shielding potential is[23]:

$$V_{\mathrm{BBC}}(\mathbf{r}) = -\sum_A \frac{Z_A - \sigma_v}{|\mathbf{r} - \mathbf{R}_A|} e^{-\sigma_v |\mathbf{r} - \mathbf{R}_A| / a_0} \tag{5}$$

Among them, $Z_A$ is the number of nuclear charges of atom A; $\sigma_v$ is the BBC blocking factor; $a_0$ is the Bohr radius; $\mathbf{R}_A$ is the position vector of atom A. We can further make the shielding parameter an adaptive function, determined by the local chemical environment of the atom and output by a machine learning model(See Supplemental Material). The ML-BBC model is constructed as:

$$V_{\mathrm{ML\text{-}BBC}}(\mathbf{r}) = -\sum_A \frac{Z_A - \sigma_v(D_A)}{|\mathbf{r} - \mathbf{R}_A|} e^{-\sigma_v(D_A) |\mathbf{r} - \mathbf{R}_A| / a_0}$$

The shielding parameter $\sigma_v$ is defined as a function of the local environment descriptor (denoted $D_A$) rather than a fixed constant.

Considering the core electronic shielding $\sigma_{core}$ and shielding attenuation length $\lambda_{screening}$, further correction is made to obtain:

$$V_{\mathrm{BBC}}(\mathbf{R}) = \frac{k_{\mathrm{Coulomb}}\left(Z - \sigma_{\mathrm{core}} - \sigma_v'\right) e^{-R/\lambda_{\mathrm{screening}}}}{R} \tag{6}$$

Where $\sigma_v'$ is the shielding parameter, and $\sigma_{\mathrm{core}} = 2.30$ is the core electron shielding constant. In the original BBC model, the exponential decay factor appears in both the numerator and denominator indices. We make a correction by separating the core-electron shielding and considering the complete shielding of the nuclear charge by inner layer electrons $\sigma_{\mathrm{core}} = 2\sigma_{1s} + 2\sigma_{2s} = 2\times 0.30 + 2\times 0.85 = 2.30$ , where $\sigma_{\mathrm{core}}$ comes from Slater's rule. We introduce a feature masking length and use an independent masking length parameter , $\lambda_{\mathrm{screening}} = 2.0\mathring{A}$ . Further, shielding attenuation length accounts for wave-function overlap, and an exponential attenuation term is introduced in the jump integration. Additionally, when $R \to 0$ , the potential energy does not diverge; When $R \to \infty$, the potential energy decays exponentially to zero; When $\sigma_v' = Z - \sigma_{core}$, the potential energy is zero, indicating complete shielding.

Graphene is a two-dimensional hexagonal lattice, with a primitive cell containing two carbon atoms. The definition of a real space basis vector is:

$$\mathbf{a}_1 = (\sqrt{3}a, 0), \quad \mathbf{a}_2 = \left( \frac{\sqrt{3}a}{2}, \frac{3a}{2} \right) \tag{7}$$

Among them, $a$ = 1.42 Å is the carbon-carbon bond length. The atomic coordinates inside the primitive cell are:

$$\mathbf{b}_1 = (0,0), \quad \mathbf{b}_2 = (0,a) \tag{8}$$

The inverted lattice basis vector is calculated using the standard relationship $\mathbf{b}_i \cdot \mathbf{a}_j = 2\pi\delta_{ij}$. The inverted lattice basis vector is calculated through standard relations.

Considering the BBC blocking modulation, the jump integral $t_{ij}$ adopts an exponential decay form:

$$\begin{cases} t_{ij}(R,\sigma_v) = t_0 \exp\left( -\dfrac{R-a}{\lambda_{\text{decay}}} \right) \left[ 1 + \alpha_t f_{\text{BBC}}(R,\sigma_v) \right] \\ f_{\text{BBC}}(R,\sigma_v) = \tanh\left( \dfrac{\delta V_{\text{BBC}}(R,\sigma_v)}{k_{\text{coulomb}} Z / a} \right) \\ \delta V_{\text{BBC}} = V_{\text{BBC}}(R,\sigma_v = 0) - V_{\text{BBC}}(R,\sigma_v) \end{cases} \tag{9}$$

Among them, $t_0 = -2.7\text{eV}$ is the unshielded nearest neighbor jump integral; $\lambda_{\text{decay}} = 0.5\text{Å}$ is the decay length of the jump integral,The function tanh ensures that the modulation factor is bounded by[-1, 1]. a=1.42 Å is carbon-carbon bond length. $\alpha_t$ is the jumping integral modulation intensity.

In situ energy $E_{onsite}$, including Coulomb self-energy, exchange correlation energy, and shielding-related linear correction:

$$E_{\text{onsite}}(\sigma_v) = E_{\text{base}} + \beta\sigma_v + \gamma f_{\text{BBC}}(R_{\text{covalent}}, \sigma_v) - \eta\sigma_v \tag{10}$$

Among them, we calculate graphene $r_{\text{covalent}} = 0.77\text{Å}$ as the covalent radius of carbon atoms; $\eta = 0.1\text{eV}$ is the exchange energy coefficient and $E_{\text{base}} = 0.0\text{eV}$ is the reference potential energy; $\beta = 0.2\text{eV}$ is the shielding linear correction coefficient. $\gamma$ is BBC potential energy modulation intensity.

Considering the non-orthogonality of the wave function, the overlapping matrix elements are constructed as follows:

$$S_{ij}(R,\sigma_v) = S_0 \exp\left(-\frac{R}{\lambda_{\text{overlap}}}\right)\left[1+\alpha_s f_{\text{BBC}}\left(R,\sigma_v\right)\right] e^{i\mathbf{k}\cdot\mathbf{R}_{ij}} \tag{11}$$

Among them, $s_0 = 0.2$ is the overlapping integration factor; $\lambda_{\text{overlap}} = 1.0\text{Å}$ is the overlapping integral attenuation length. $\alpha_s$ is the overlapping integral modulation intensity.。

In inverse space, the Hamiltonian matrix elements are:

$$H_{ij}(\mathbf{k}) = \sum_{\mathbf{R}} t_{ij}(R) e^{i\mathbf{k}\cdot\mathbf{R}} + \delta_{ij} E_{\text{onsite}}(\sigma_v) \tag{12}$$

The corresponding overlap matrix is:

$$S_{ij}(\mathbf{k}) = \delta_{ij} + \sum_{\mathbf{R}} S_{ij}(R) e^{i\mathbf{k}\cdot\mathbf{R}} \tag{13}$$

The intrinsic energy spectrum is obtained by solving the generalized eigenvalue problem:

$$H(\mathbf{k})\psi_n(\mathbf{k}) = E_n(\mathbf{k}) S(\mathbf{k})\psi_n(\mathbf{k}) \tag{14}$$

Density of states (DOS) is calculated by integrating the entire Brillouin zone:

$$\text{DOS}(E) = \frac{1}{N_k N_{\text{atom}}} \sum_{\mathbf{k}} \sum_{n} \frac{1}{\sigma\sqrt{2\pi}} \exp\left[-\frac{\left(E - E_n(\mathbf{k})\right)^2}{2\sigma^2}\right] \tag{15}$$

Among them, $N_k$ is the total number of *k* points; $N_{\text{atom}}$ is the number of atoms in the original cell; $\sigma = 0.05\text{eV}$ is the Gaussian broadening width.

## 2.2 Spiral wave function

In momentum space, the Bloch Hamiltonian can be expressed as:

$$\hat{H}(\mathbf{k}) = \begin{pmatrix} 0 & -tf(\mathbf{k}) \\ -tf^*(\mathbf{k}) & 0 \end{pmatrix} \tag{16}$$

Among them, the structural factor $f(\mathbf{k})$ is defined as:

$$f(\mathbf{k}) = \sum_{i=1}^{3} e^{i\mathbf{k}\cdot\boldsymbol{\delta}_i} = e^{i\mathbf{k}\cdot\boldsymbol{\delta}_1} + e^{i\mathbf{k}\cdot\boldsymbol{\delta}_2} + e^{i\mathbf{k}\cdot\boldsymbol{\delta}_3} \tag{17}$$

By substituting the specific form of the real space basis vector, we can simplify it into:

$$f(\mathbf{k}) = e^{ik_y a/\sqrt{3}} + 2e^{-ik_y a/(2\sqrt{3})} \cos\left(\frac{k_x a}{2}\right) \tag{18}$$

The position of the Dirac point and low-energy expansion indicate that the Dirac point appears at the position of $f(\mathbf{k})=0$, resulting in two non-equivalent Dirac points:

$$\mathbf{K}=\left(\frac{4\pi}{3a},0\right),\quad \mathbf{K}'=\left(\frac{2\pi}{3a},\frac{2\pi}{\sqrt{3}a}\right) \tag{19}$$

Near point *K*, let $k=K+q$, where $|q|$ is very small. Expanding the first order from $f\left(k\right)$ to $q$:

$$f(\mathbf{K}+\mathbf{q})=e^{iq_y a/\sqrt{3}}+2e^{-iq_y a/(2\sqrt{3})}\cos\left(\frac{K_x a+q_x a}{2}\right) \tag{20}$$

Substituting $K_x=4\pi/3a$:

$$f(\mathbf{K}+\mathbf{q})=e^{iq_y a/\sqrt{3}}+2e^{-iq_y a/(2\sqrt{3})}\cos\left(\frac{2\pi}{3}+\frac{q_x a}{2}\right) \tag{21}$$

Using the trigonometric identity:

$$\cos\left(\frac{2\pi}{3}+\theta\right)=-\frac{1}{2}\cos\theta-\frac{\sqrt{3}}{2}\sin\theta \tag{22}$$

And performing a first-order expansion on small $q$:

$$\begin{cases} e^{iq_y a/\sqrt{3}}\approx 1+iq_y a/\sqrt{3} \\ e^{-iq_y a/(2\sqrt{3})}\approx 1-iq_y a/(2\sqrt{3}) \\ \cos(q_x a/2)\approx 1 \\ \sin(q_x a/2)\approx q_x a/2 \end{cases} \tag{23}$$

Substituting and retaining the first-order term:

$$\begin{aligned} f(\mathbf{K}+\mathbf{q}) &\approx \left(1+iq_y a/\sqrt{3}\right)+2\left(1-iq_y a/(2\sqrt{3})\right)\left(-\frac{1}{2}-\frac{\sqrt{3}}{2}\frac{q_x a}{2}\right) \\ &=\left(1+iq_y a/\sqrt{3}\right)+\left(1-iq_y a/(2\sqrt{3})\right)\left(-1-\sqrt{3}\frac{q_x a}{2}\right) \\ &\approx \frac{\sqrt{3}a}{2}\left(iq_y-q_x\right)=-\frac{\sqrt{3}a}{2}\left(q_x-iq_y\right) \end{aligned} \tag{24}$$

Substituting the expanded result into the Hamiltonian:

$$\hat{H}(\mathbf{K}+\mathbf{q})\approx\begin{pmatrix} 0 & t\frac{\sqrt{3}a}{2}(q_x-iq_y) \\ t\frac{\sqrt{3}a}{2}(q_x+iq_y) & 0 \end{pmatrix} \tag{25}$$

If the Fermi velocity $v_F = \frac{\sqrt{3}at}{2\hbar}$ is defined, then:

$$\hat{H}_{\mathbf{K}}(\mathbf{q}) \approx \hbar v_F \begin{pmatrix} 0 & q_x - iq_y \\ q_x + iq_y & 0 \end{pmatrix} = \hbar v_F \left( q_x\sigma_x + q_y\sigma_y \right) \tag{26}$$

Where $\sigma_x$ and $\sigma_y$ is the Pauli matrix.

The eigenvalue equation of the Hamiltonian $H = \hbar v_F q \cdot \sigma$ is:

$$H\psi = E\psi \tag{27}$$

Solving for the eigenvalues:

$$E_\lambda(\mathbf{q}) = \lambda \hbar v_F |\mathbf{q}|, \quad \lambda = \pm 1 \tag{28}$$

Where $\lambda = +1$ corresponds to the conduction band, and $\lambda = -1$ corresponds to the valence band.

Intrinsic state, normalized wave function is given as:

$$\psi_{\lambda,\mathbf{K}}(\mathbf{q}) = \frac{1}{\sqrt{2}} \begin{pmatrix} 1 \\ \lambda e^{i\phi_{\mathbf{q}}} \end{pmatrix} \tag{29}$$

Where $\phi_{\mathbf{q}} = \arctan\left( q_y / q_x \right)$ is the azimuth of momentum $q$.

For $\lambda = +1$ (conduction band):

$$\begin{aligned} H\psi &= \frac{\hbar v_F}{\sqrt{2}} \begin{pmatrix} 0 & q_x - iq_y \\ q_x + iq_y & 0 \end{pmatrix} \begin{pmatrix} 1 \\ e^{i\phi_{\mathbf{q}}} \end{pmatrix} \\ &= \frac{\hbar v_F}{\sqrt{2}} \begin{pmatrix} (q_x - iq_y)e^{i\phi_{\mathbf{q}}} \\ q_x + iq_y \end{pmatrix} \\ &= \frac{\hbar v_F |\mathbf{q}|}{\sqrt{2}} \begin{pmatrix} 1 \\ e^{i\phi_{\mathbf{q}}} \end{pmatrix} = E\psi \end{aligned} \tag{30}$$

For point $K'$, similar expansion can lead to:

$$\hat{H}_{\mathbf{K}'}(\mathbf{q}) \approx \hbar v_F \begin{pmatrix} 0 & q_x + iq_y \\ q_x - iq_y & 0 \end{pmatrix} = \hbar v_F \left( -q_x\sigma_x + q_y\sigma_y \right) \tag{31}$$

The corresponding wave function:

$$\psi_{\lambda,\mathbf{K}'}(\mathbf{q}) = \frac{1}{\sqrt{2}} \begin{pmatrix} 1 \\ -\lambda e^{-i\phi_{\mathbf{q}}} \end{pmatrix} \tag{32}$$

Helicity (chirality) is dominated by the phase factor in the wave function, and its core characteristics include[18, 24, 25]:

1. Momentum space phase vortex: Around the Dirac point, the wave function undergoes a phase change of $2\pi$.

2. Pseudo spin momentum locking: Calculate the expected value of pseudo spin:

$$\langle\sigma_x\rangle = \lambda\cos\phi_{\mathbf{q}},\quad \langle\sigma_y\rangle = \lambda\sin\phi_{\mathbf{q}},\quad \langle\sigma_z\rangle = 0$$

Display pseudo spin and momentum locking, where the pseudo spin is in the momentum plane and oriented parallel (conduction band) or anti-parallel (valence band) to the momentum direction.

3. Non-trivial topology and Berry phase: The Berry phase of adiabatic motion around a Dirac point is $\gamma = \oint \mathbf{A}(\mathbf{q})\cdot d\mathbf{q} = \pi$ (Berry connection $\mathbf{A}(\mathbf{q}) = i\langle\psi|\nabla_{\mathbf{q}}|\psi\rangle$);

4. Chiral symmetry: The chirality of point $K$ is opposite to that of point $K'$.

## 3. Results and Discussion

### 3.1 Electrostatic shielding potential

**Figure 1** presents the evolution of BBC shielding potential and jump integral with interatomic distance (0.5–5.0 Å), and compares the differences in physical properties under different shielding parameters $\sigma_v$. The data show small but distinct differences in both types of physical quantities, indicating that regulation of shielding parameters can quantifiably affect strength of electron interactions. The improved BBC potential ranges from -82.98 eV to 0.00 eV, while the jump integral varies from -17.00 eV to 0 eV, reflecting significant differences in the spatial scale of different physical quantities.

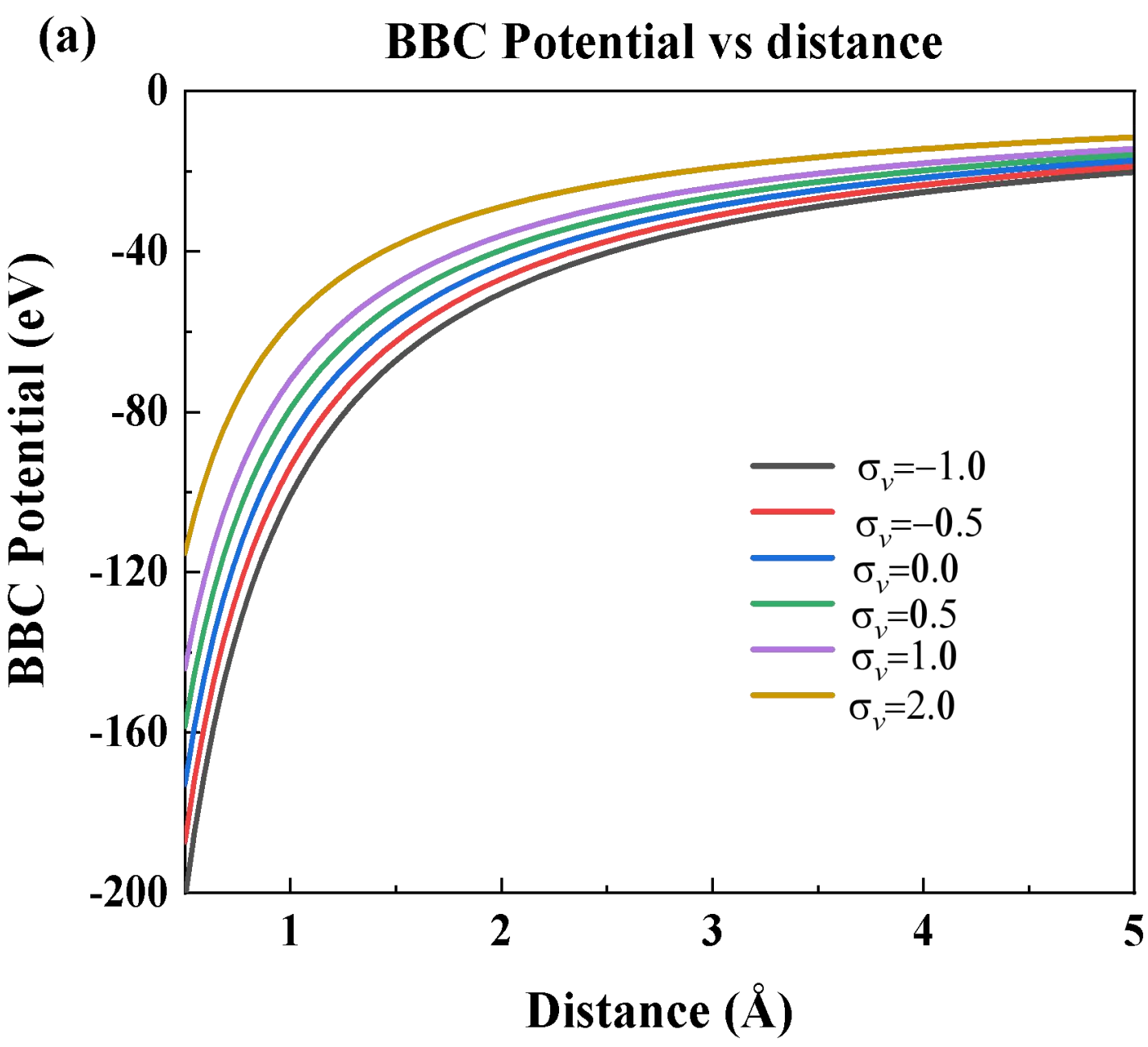

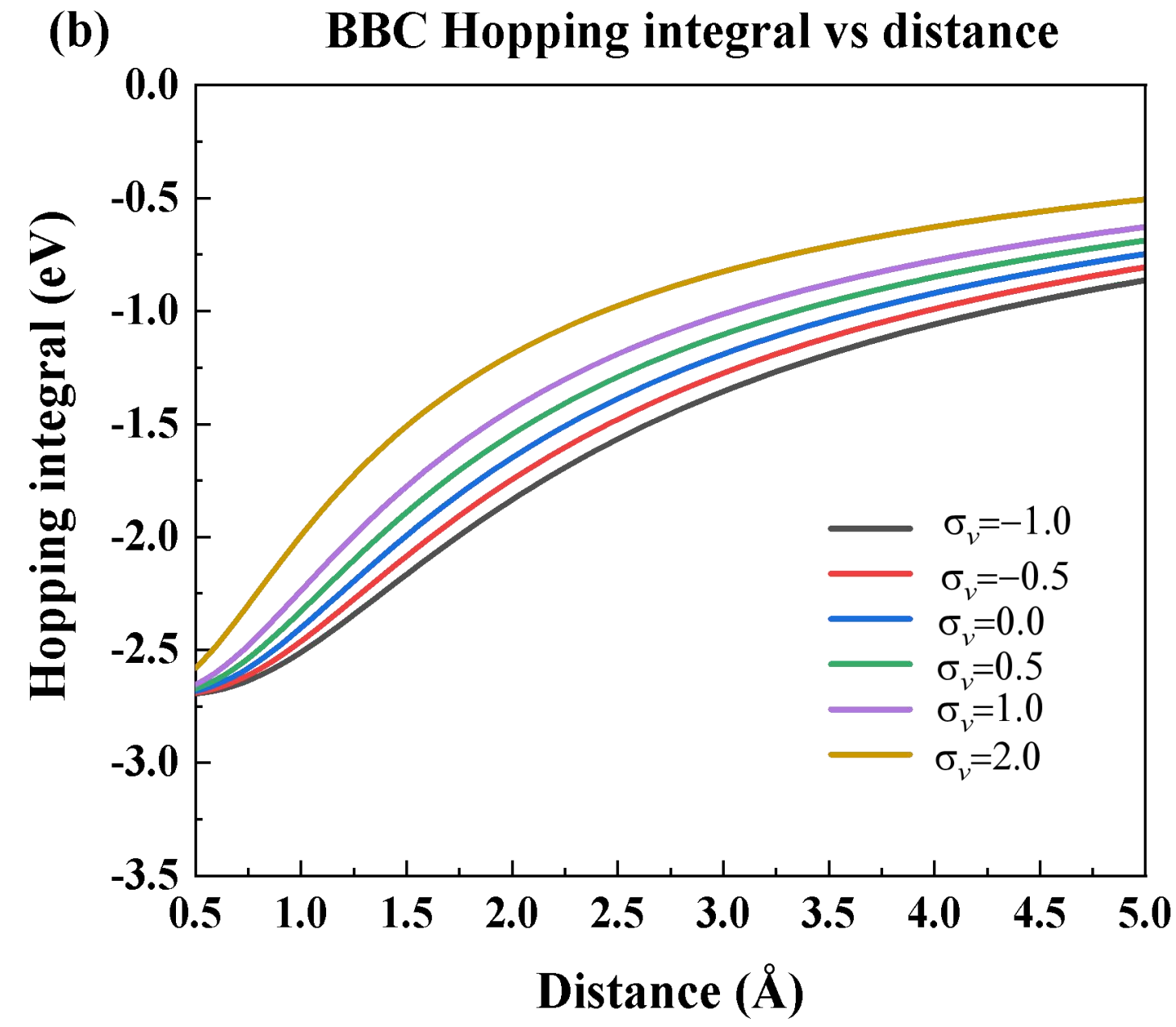

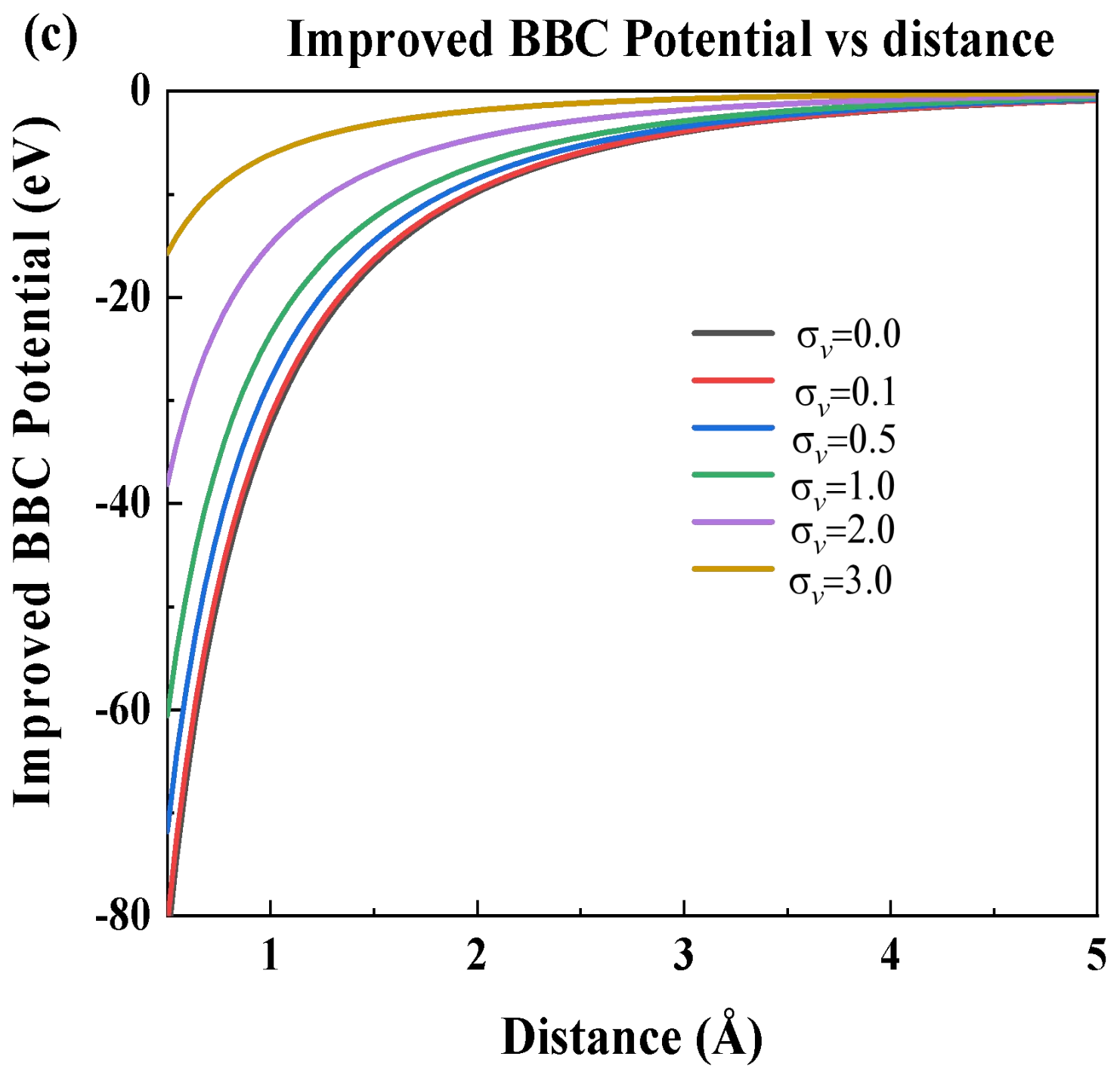

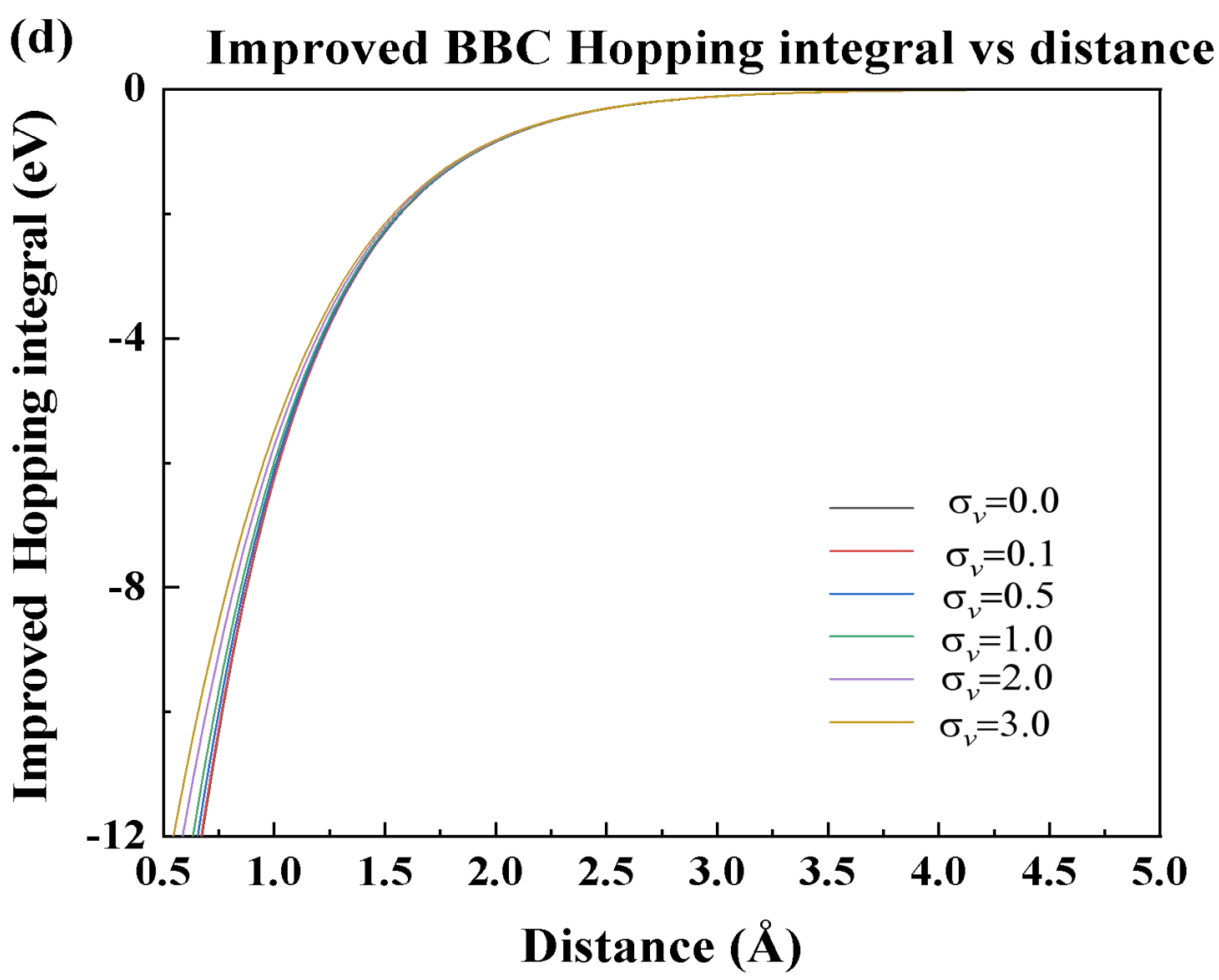

**Figure 1** Comparison of BBC Potential Energy and Improved BBC Potential Energy with Atomic distance at Different Shielding Parameters $\sigma_v$

The BBC potential exhibits typical shielding Coulomb potential characteristics, rapidly decaying with increasing interatomic distance. Near the nearest neighbor atomic distance 1.42Å of graphene, the Hopping integral energy is about 1.94 eV, indicating a strong shielding effect. The decay behavior conforms to the mathematical form of exp (- $R/\lambda$)/$R$. The slight difference between the two data types reflects the modulation effect of the shielding parameter $\sigma_v$ on the effective charge ($Z - \sigma_{core} - \sigma_v$).

The jump integral in the improved model has a negative value (corresponding to bonding interactions), and its absolute value gradually decreases with the increase of interatomic distance. Data indicate that within the distance range of 0.5 Å to 5.0 Å, the absolute value of the jump integral decreased by about 100%. This phenomenon results from the exponential decay of the overlap degree of the electronic wave function with increasing distance, highlighting the sensitivity of the transfer integral to atomic spacing in the tight-binding approximation.

Comparing two types of physical quantities provides a direct theoretical basis for the quantitative study of shielding effects. A small change in the shielding parameter $\sigma_v$ can trigger a significant change in the jump integral at a specific distanc. This sensitivity suggests that altering the shielding environment via chemical doping or gate voltage regulation significantly alters the electronic structure of graphene. Additionally, the data validate the improved BBC model: the BBC potential does not diverge at short distances, effectively avoiding the singularity problem of the bare Coulomb potential, and the jump integral approaches zero at long distances, aligning well with actual physical situations. In experimental research, localized in-situ energy is accurately measured using characterization methods such as X-ray photoelectron spectroscopy (XPS). Comparing theoretical calculations with experimental data is essential for calibrating shielding parameters.

### 3.2 Band structure and density of states

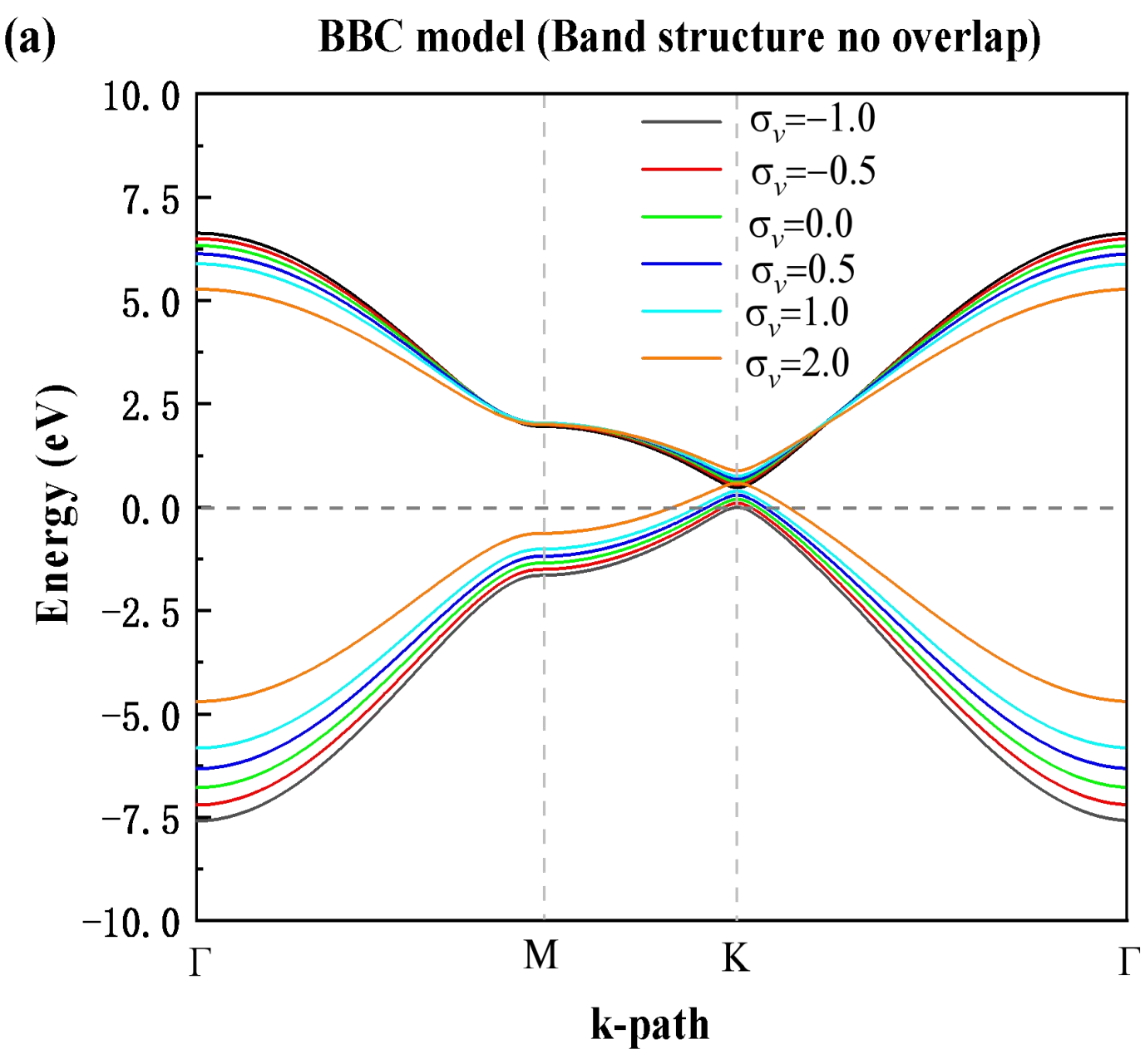

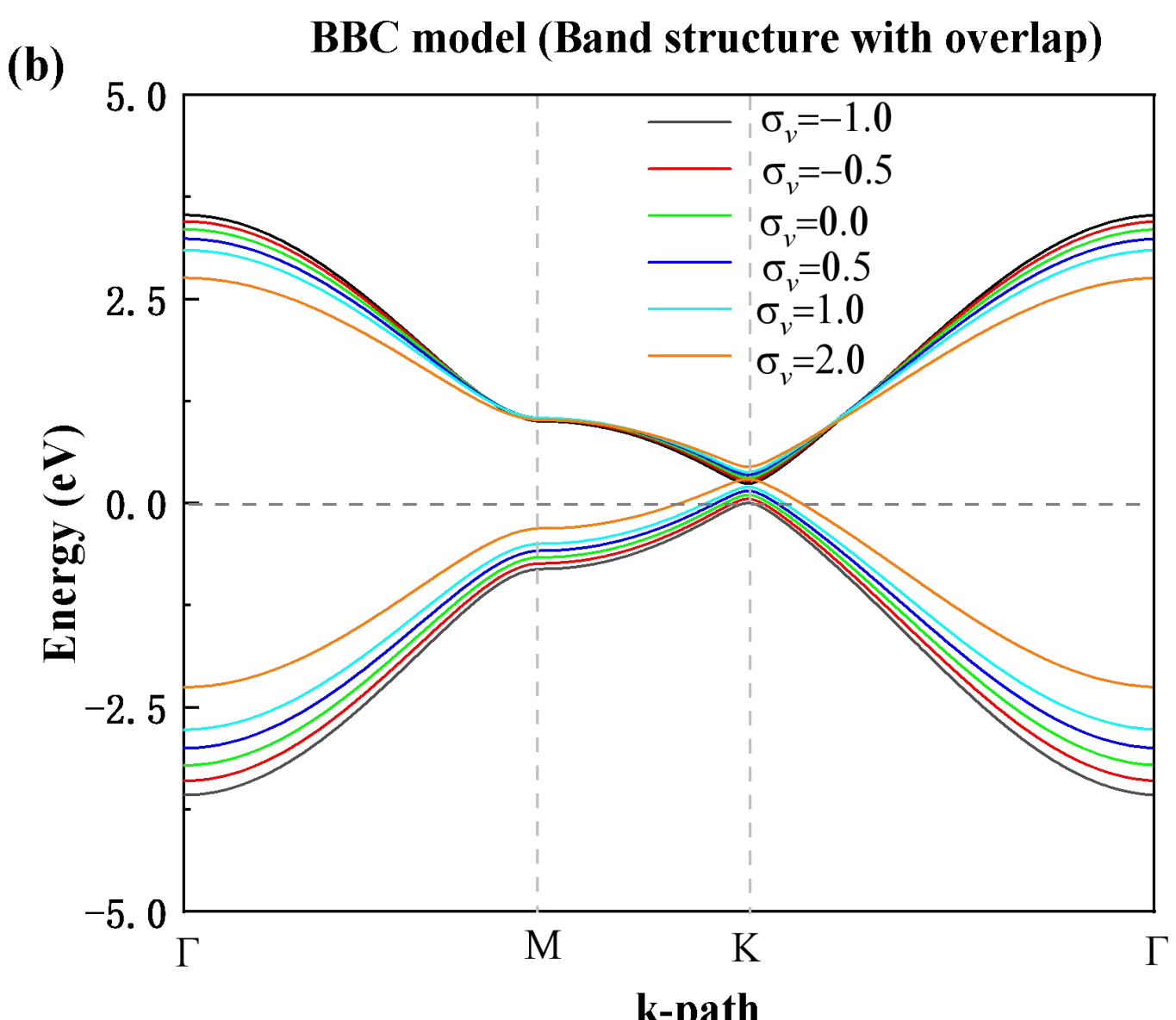

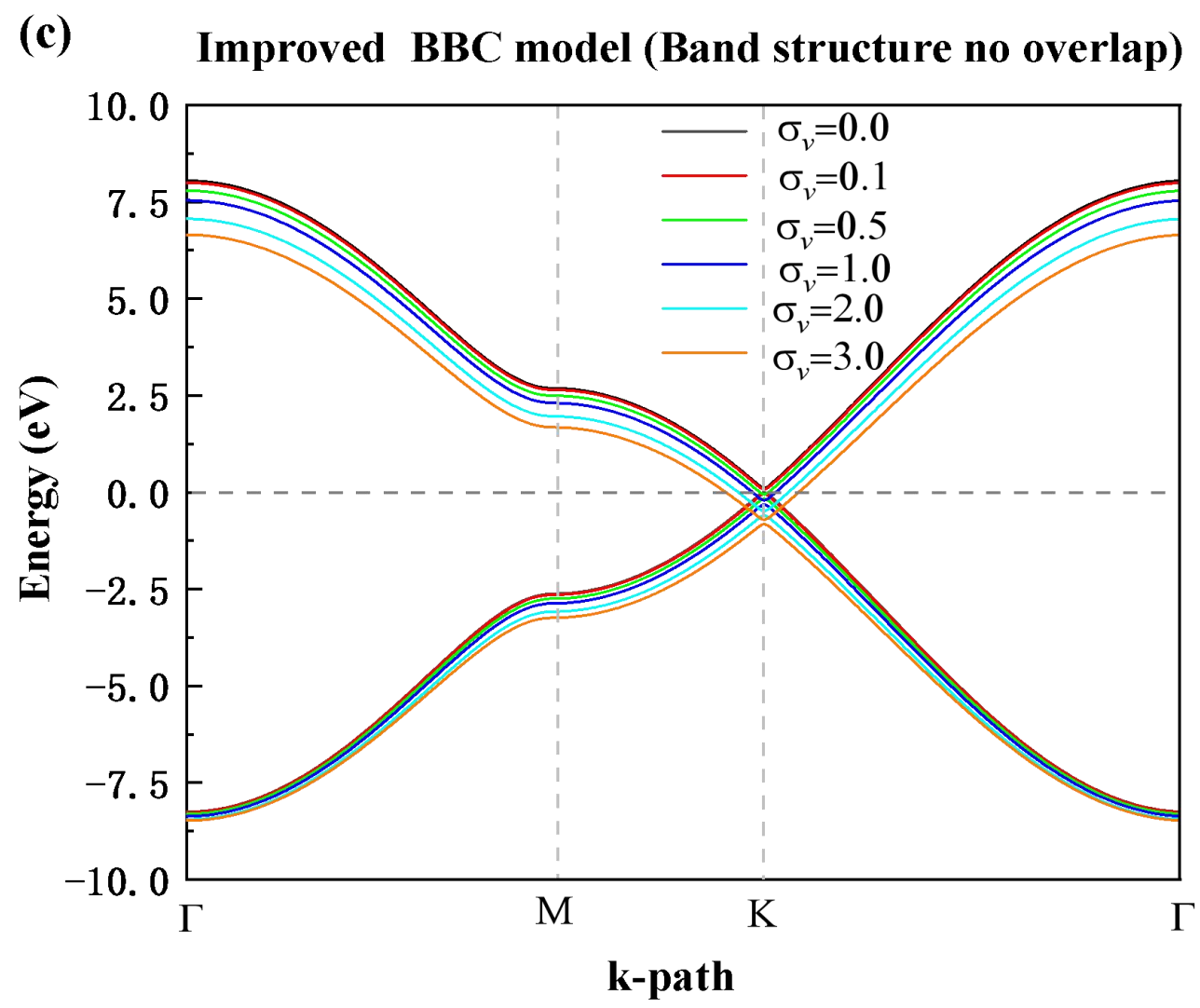

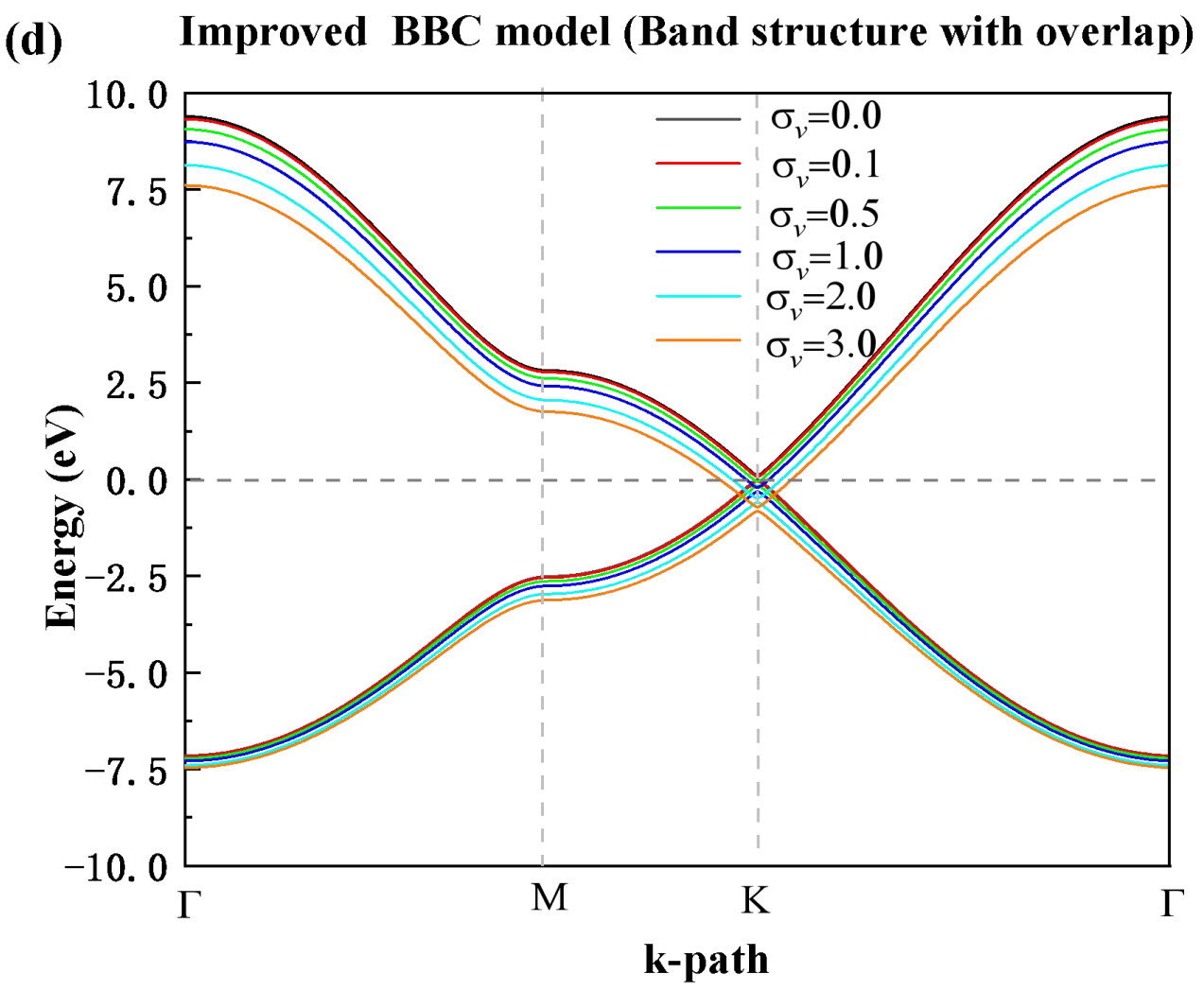

**Figure 2** Band structure of graphene under different shielding parameters: (a) BBC model (non overlapping integration); (b) BBC model (including overlapping integral); (c) Improved BBC model (non overlapping integral); (d) Improved BBC model (including overlapping integral)

**Figure 2 (a)** presents the band energy evolution characteristics of graphene along the highly symmetric path of Brillouin zone (Γ→M→K→Γ) under different shielding parameters $\sigma_v$ . As shielding strength $\sigma_v$ increases, the energy of all high symmetry points (Γ, M, K) exhibits a systematic downward trend, causing the entire band structure to shift to lower energy. A comparison of the band structures between the BBC Model and DFT calculations shows consistency.[26]**(See Supplemental Material)**Notably, different *k* points respond differently to shielding effects, with the K point showing the largest energy reduction. This indicates that the shielding effect weakens electron interaction and leads to band restructuring, selectively affecting different electronic states in momentum space. **Figure 2(b)** presents the band calculation results after introducing overlap integration in the BBC model. Compared to **Figure 2(a)**, the introduction of the overlap integral significantly narrows the band width, reflecting the modulation effect of orbital overlap on the distribution of electronic states.

**Figure 2(c)**, based on an improved BBC model, reveals the evolution law of graphene energy bands with shielding strength $\sigma_v$. When $\sigma_v$=0, the system is at the strong interaction limit without shielding, where the energy at point Γ is the highest and the Dirac cone at point *K* is adjacent to the Fermi level; As $\sigma_v$ gradually increases, the energy at the Γ point decreases significantly, while the energy at the *K* point shows a clear upward trend. This opposite trend indicates that the shielding effect has a significant directional dependence on electronic states at different high symmetry points in the Brillouin zone. The shielding effect does not uniformly modulate the entire energy band but selectively alters electron correlation strength in specific momentum regions, potentially impacting the system's topological properties and transport behavior.Based on the comparison of the subgraphs in **Figure 2**, it can be seen that the calculation results with and without overlapping integrals in the BBC model show consistent overall trends in band width variation with $\sigma_v$, while the introduction of overlapping integrals further exacerbates the narrowing of band width.

**Figure 2** clearly illustrates the shielding induced band rearrangement phenomenon: when $\sigma_v$ is small ($\sigma_v$ <1.00), the energy at point *K* remains stable, consistent with the linear dispersion characteristics of graphene Dirac cones; When $\sigma_v \geq 1.00$, the energy at point *K* significantly increases, indicating that a strong shielding environment may open a bandgap or alter the geometry of the Dirac cone. Meanwhile, the energy of the points Γ and M continues to decrease with increasing shielding strength, suggesting a stronger regulatory effect of the shielding on the electronic states near the center of the Brillouin zone. This phenomenon may trigger a redistribution of the effective mass of electrons.

Previous studies have demonstrated the size effect of Ag nanoparticles induced by charge shielding observed in STS and XPS spectroscopy[27, 28]. The microscopic mechanism of band renormalization arises from the spatial modulation of Coulomb interactions between electrons due to shielding effects.Under low shielding conditions (with small $\sigma_v$), long-range Coulomb interactions dominate, leading to strong localization of the electronic wave function and raising the band energy near point Γ. As shielding strength increases ($\sigma_v$ increases), the effective range of the Coulomb potential decreases, enhancing delocalization of the electron wave function, reducing energy near point Γ, and weakening the stability of the Dirac cone at point *K*, which facilitates the opening of the energy gap. This process demonstrates that the shielding effect can finely control the electronic topological properties of graphene by regulating electronic correlation strength.

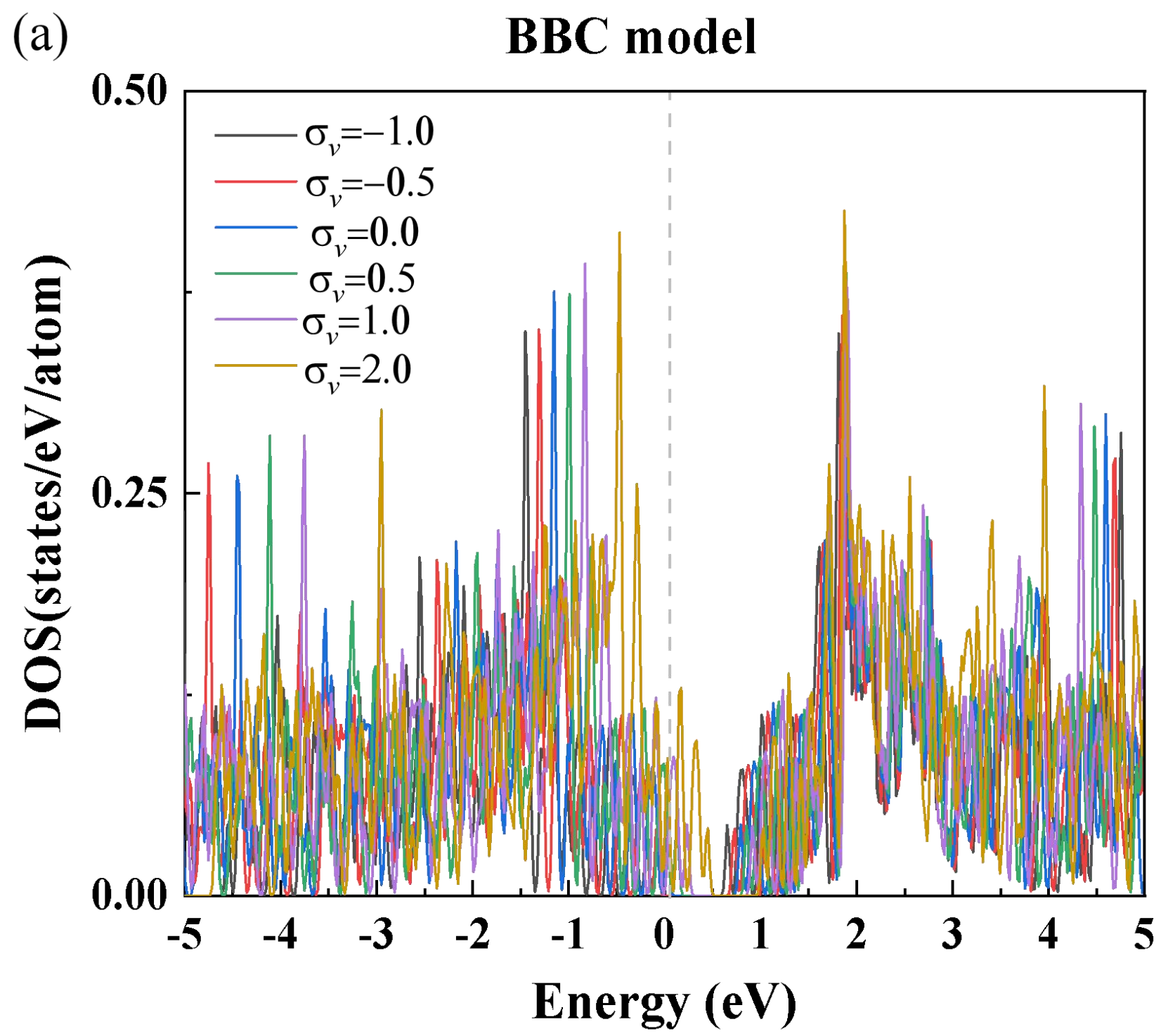

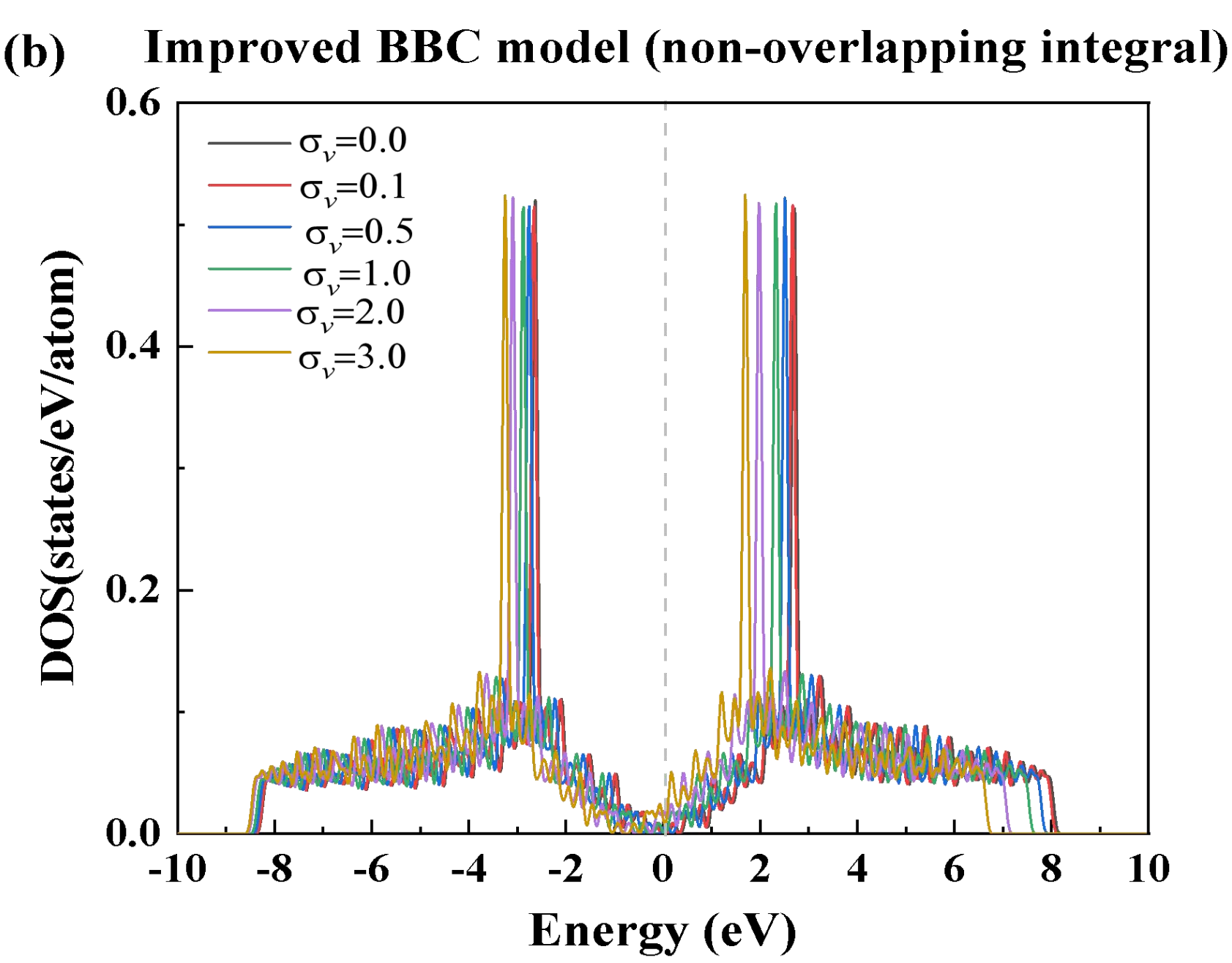

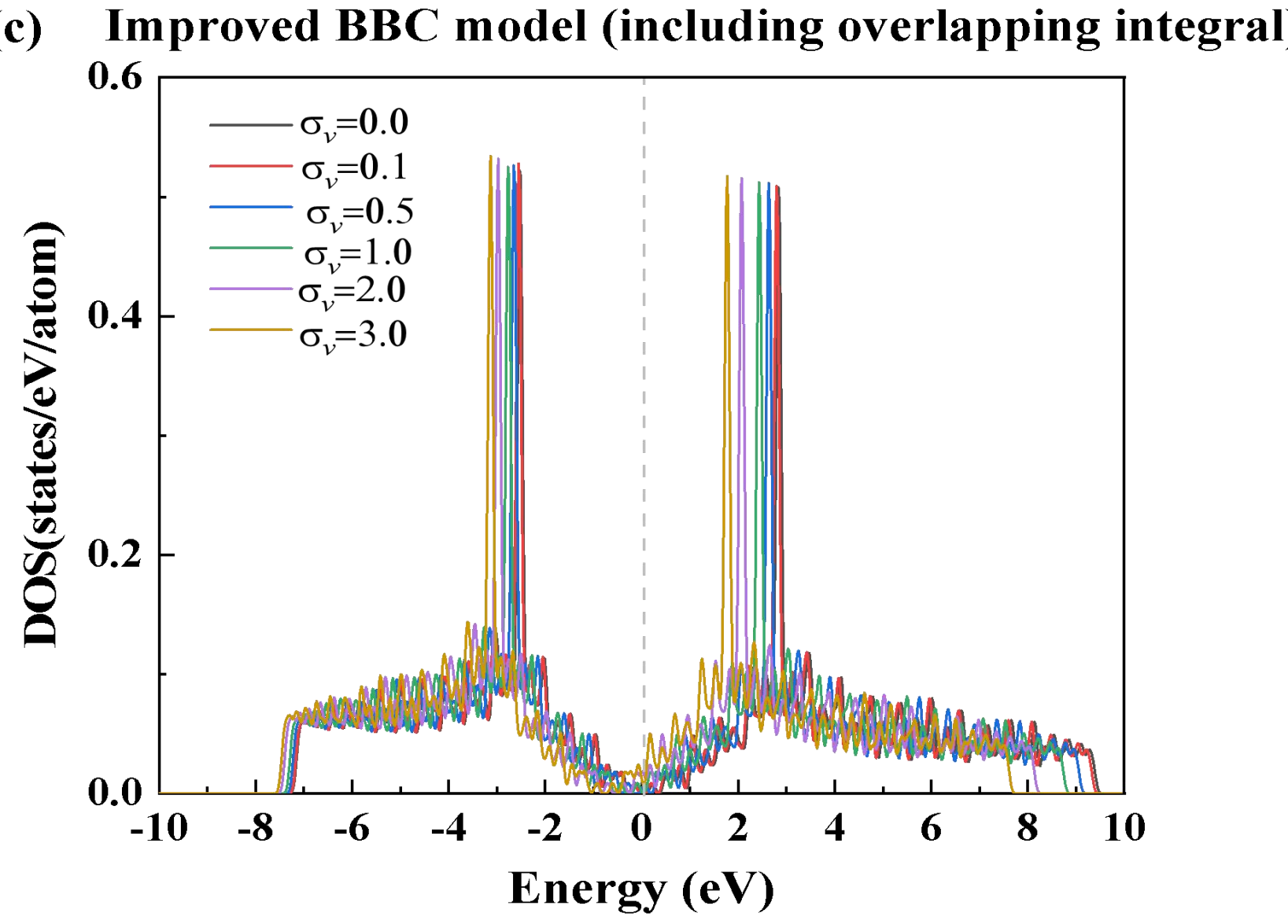

**Figure 3** Density of states of graphene under different shielding parameters: (a) BBC model ; (b) Improved BBC model (non-overlapping integral); (c) Improved BBC model (including overlapping integral)

**Figure 3(a)** shows that as the shielding parameter $\sigma_v$ increases from -1.00 to 2.00, the density of states at each energy point generally increases monotonically.The y-axis represents the DOS in units of states/eV/atom, which reflects the number of available electronic states at a given energy. As can be seen, a prominent DOS peak appears in the energy range of approximately -1 eV to 0 eV, and a significant increase in DOS is also observed between 1 eV and 3 eV. In contrast, the DOS is generally low at energies below -4 eV and above 4 eV. This provides a quantitative basis for understanding the renormalization mechanism of shielding on electronic structures.

**Figure 3(b)** presents the density of states distribution of graphene in the energy range of -10 eV to 10 eV under different shielding strengths $\sigma_v$. The data show that the density of states reaches its maximum (0.55 states/eV/atom) near the Fermi level (0 eV), consistent with the linear dispersion relation near the Dirac cone of graphene. As the energy deviates from the Fermi level, the density of states gradually decreases and approaches zero at ±10 eV, conforming to the typical behavior of two-dimensional material density of states.As can be seen, two sharp main peaks appear at approximately −2 eV and 2 eV, while relatively flat DOS plateaus exist in the ranges of −6 eV to −2 eV and 2 eV to 6 eV. In contrast, the DOS is generally low in the regions of −10 eV to −6 eV and 6 eV to 10 eV. The overall trend of the DOS curves under different strains is highly consistent, with only subtle differences in peak height and fine details, indicating that strain has a relatively limited modulation effect on the electronic structure of the material within this model. Meanwhile, the Improved BBC model (with non-overlapping integral) exhibits sharper DOS peaks and clearer structural features, reflecting an improved accuracy in describing the distribution of electronic states.This indicates that the shielding effect mainly influences high-energy electronic states by modifying the band width and changing the band curvature. **Figure 3(c)** shows the density of states of the improved BBC model with overlapping integration, and its overall trend is consistent with the case without overlapping integration (**Figure 3(b)**). The introduction of overlapping integration does not alter the dependence of density on shielding parameters.

### 3.3 Spiral wave function and Berry curvature

The low-energy quasi-particles of graphene near the Dirac point are described by the massless Dirac equation, and its intrinsic wave function (taking point $K$ as an example) has the following form:

$$\psi_{\lambda,\mathbf{K}}(\mathbf{q})=\frac{1}{\sqrt{2}}\begin{pmatrix}1\\ \lambda e^{i\phi_{\mathbf{q}}}\end{pmatrix} \tag{33}$$

Its core feature stems from the phase factor $e^{i\phi_{\mathbf{q}}}$ related to the azimuth angle in momentum space, which directly leads to the wave function having a non-trivial vortex structure around the Dirac point $(q=0)$ in momentum space. When the momentum vector $q$ rotates around the Dirac point once ($\phi_{\mathbf{q}}$ changes $2\pi$), the wave function gains a phase accumulation of $2\pi$, which is mathematically isomorphic to the orbital angular momentum vortex beam in optics, with a topological charge of $l=1$. Further physics is reflected in the pseudo-spin texture. The calculation shows that the expected value of pseudo spin (representing the degree of freedom of the lattice) under this wave function is:

$$\langle\sigma_x\rangle=\lambda\cos\phi_{\mathbf{q}},\quad\langle\sigma_y\rangle=\lambda\sin\phi_{\mathbf{q}} \tag{34}$$

This means that the pseudo spin vector is strictly located in the plane and its direction is locked to the momentum direction $\phi_{\mathbf{q}}$ : parallel in the conduction band ($\lambda=+1$) and anti-parallel in the valence band ($\lambda=-1$). This pseudo-spin-momentum locking relationship is the essence of graphene's electronic chirality, which determines its unique scattering properties: it can effectively suppress backscattering at one-dimensional potential barriers, a quantum-mechanical root of graphene's high carrier mobility.

Our derivation further elucidates that at another non-equivalent Dirac point ($K'$ point), the low-energy effective Hamiltonian and wave function can be expressed as:

$$\hat{H}_{\mathbf{K}'}(\mathbf{q}) \approx \hbar v_F \left(-\sigma_x q_x + \sigma_y q_y\right), \quad \psi_{\lambda,\mathbf{K}'}(\mathbf{q}) = \frac{1}{\sqrt{2}}\begin{pmatrix} 1 \\ -\lambda e^{-i\phi_{\mathbf{q}}} \end{pmatrix} \tag{35}$$

Compared with the point, its helical phase has opposite chirality ( $e^{-i\phi_{\mathbf{q}}}$ ), topological charge is -1, and the vortex direction of the pseudo-spin texture is also opposite. The K point and K' point together constitute the Valley Degree of Freedom of graphene, and these two valleys are interrelated under time reversal symmetry. The opposite chirality between valleys lays the theoretical foundation for selectively manipulating electrons in specific valleys through electric fields, strain, or boundary control, known as “valley electronics” research.

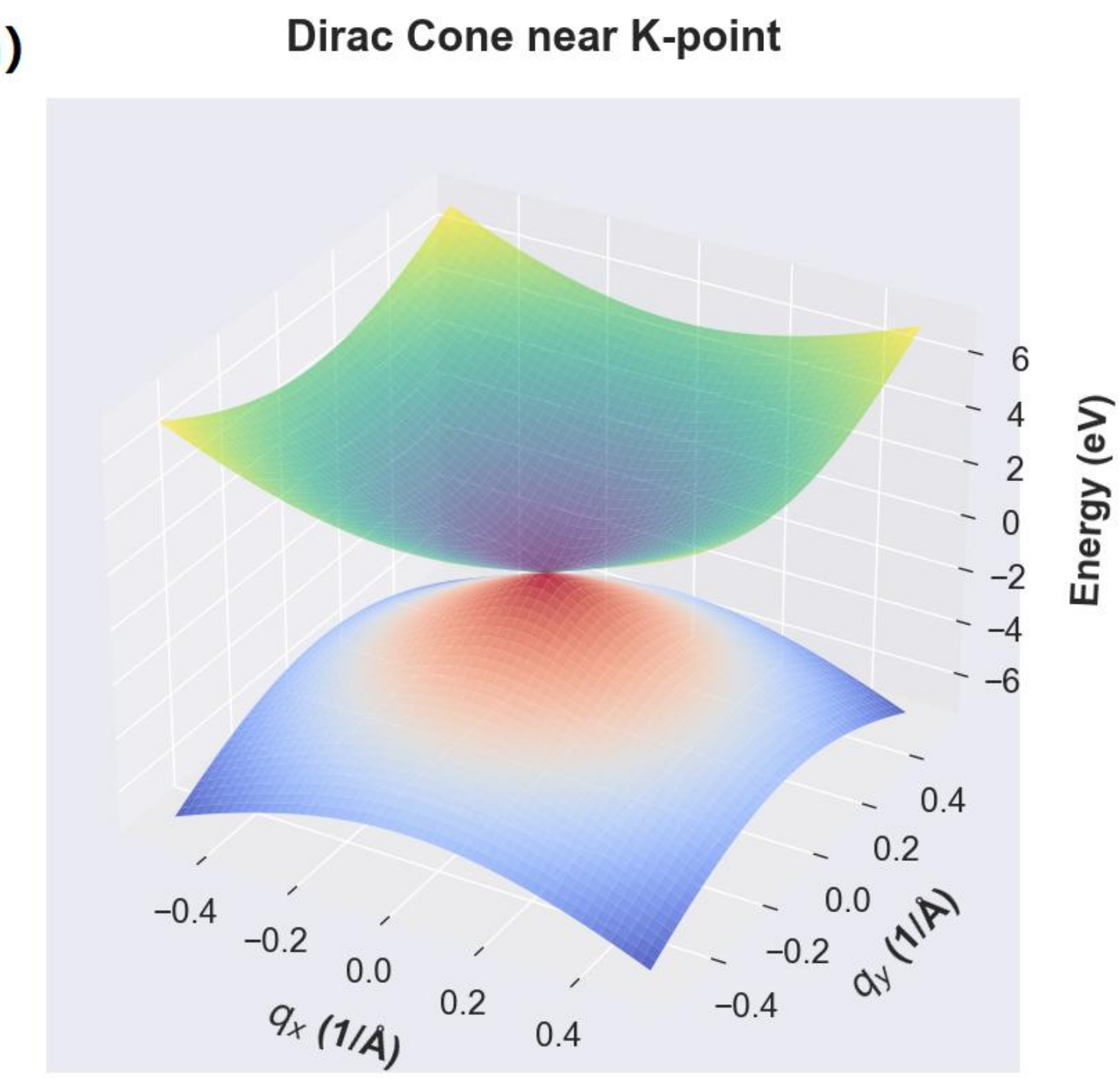

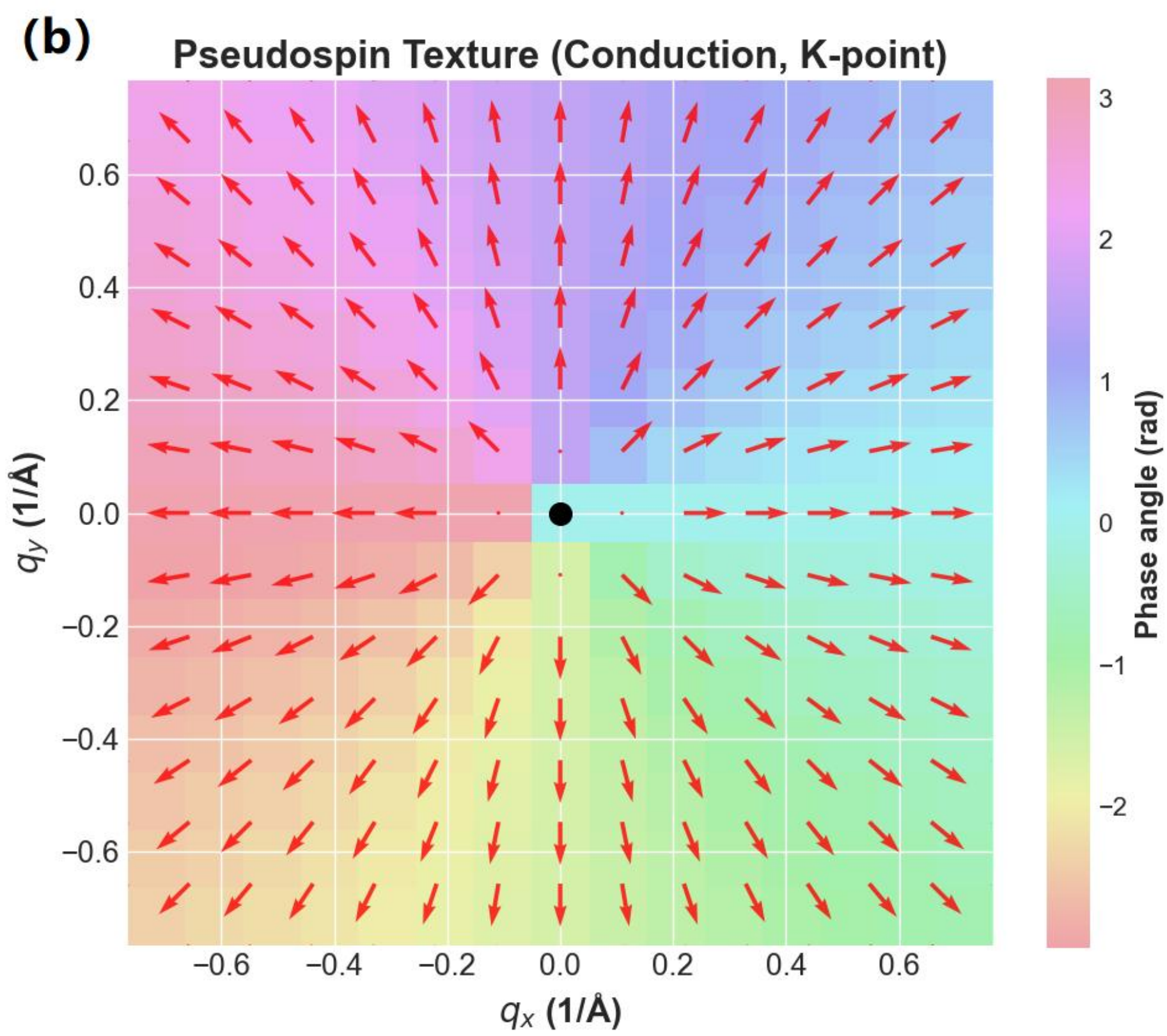

**Figure 4** (a) Dirac cone near point *K* on a three-dimensional surface; (b) Pseudo spin texture in momentum space

**Figure 4(a)** depicts the Dirac cone near point *K* in three-dimensional curved form, visually demonstrating the linear dependence of energy and momentum, and the cone surface is symmetrical vertically. **Figure 4(b)** shows the pseudospin texture in momentum space: the direction of the arrow represents the expected direction of pseudospin (pseudovector of subgrid degrees of freedom) in the conduction band. The pseudo spin is completely locked in the direction of momentum; on any circumference around the Dirac point, the pseudo spin is arranged along the tangential direction, forming a planar vortex. This texture directly corresponds to the phase factor $e^{i\phi_{\mathbf{q}}}$ of the spiral wave function, which is a visual manifestation of the chirality of graphene electrons.

A direct and significant consequence of the spiral wave function is the generation of non-zero Berry phases. When electrons undergo adiabatic evolution along a closed path around the Dirac point in momentum space, their wave function accumulates a Berry phase $\gamma$ determined by the system geometry in addition to the dynamic phase. For the above wave function, $\gamma = \pi$ can be calculated.

This phase is a key indicator of the non-mediocrity of graphene band topology. It profoundly affects the quantum transport behavior of graphene:

1. Half-integer quantum Hall effect: Under strong magnetic fields, traditional two-dimensional electron gases exhibit integer quantum Hall effect. Due to the existence of the $\pi$ Berry phase in graphene, the degeneracy and sequence of its Landau energy levels change, ultimately leading to the appearance of quantum Hall platforms at fill factors of ν=± 2, ± 6, ± 10,..., where the platform at ν=±2 is a characteristic manifestation of its "semi-integer" shift.

2. Suppression of weak anti-localization: In disordered systems, the $\pi$ Berry phase causes the quantum interference of time reversal path pairs to shift from destructive (weak localization) to constructive (weak anti-localization), resulting in a characteristic positive magnetic resistance signal in low magnetic field magnetoresistance. This theory is perfectly consistent with experimental observations.

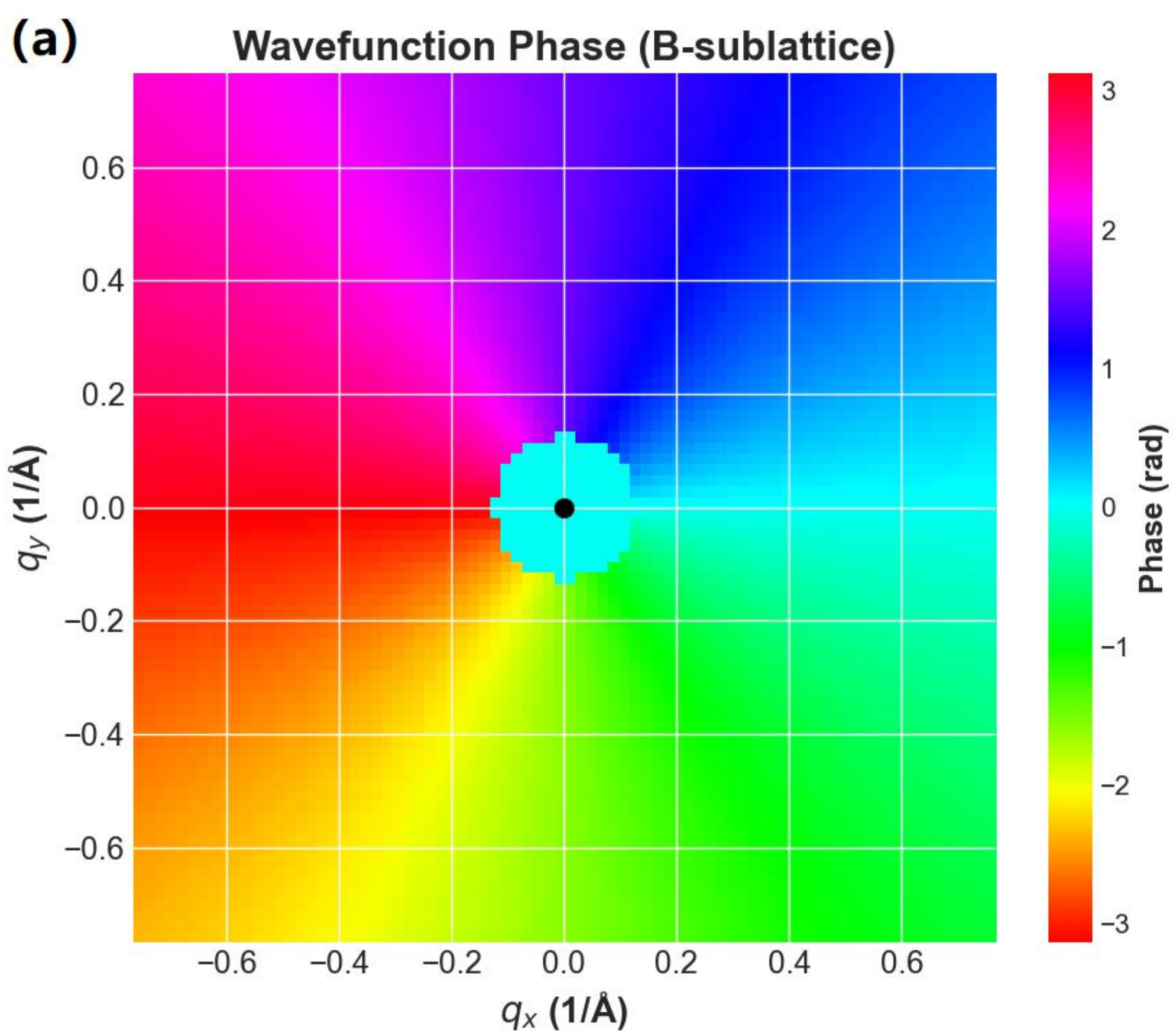

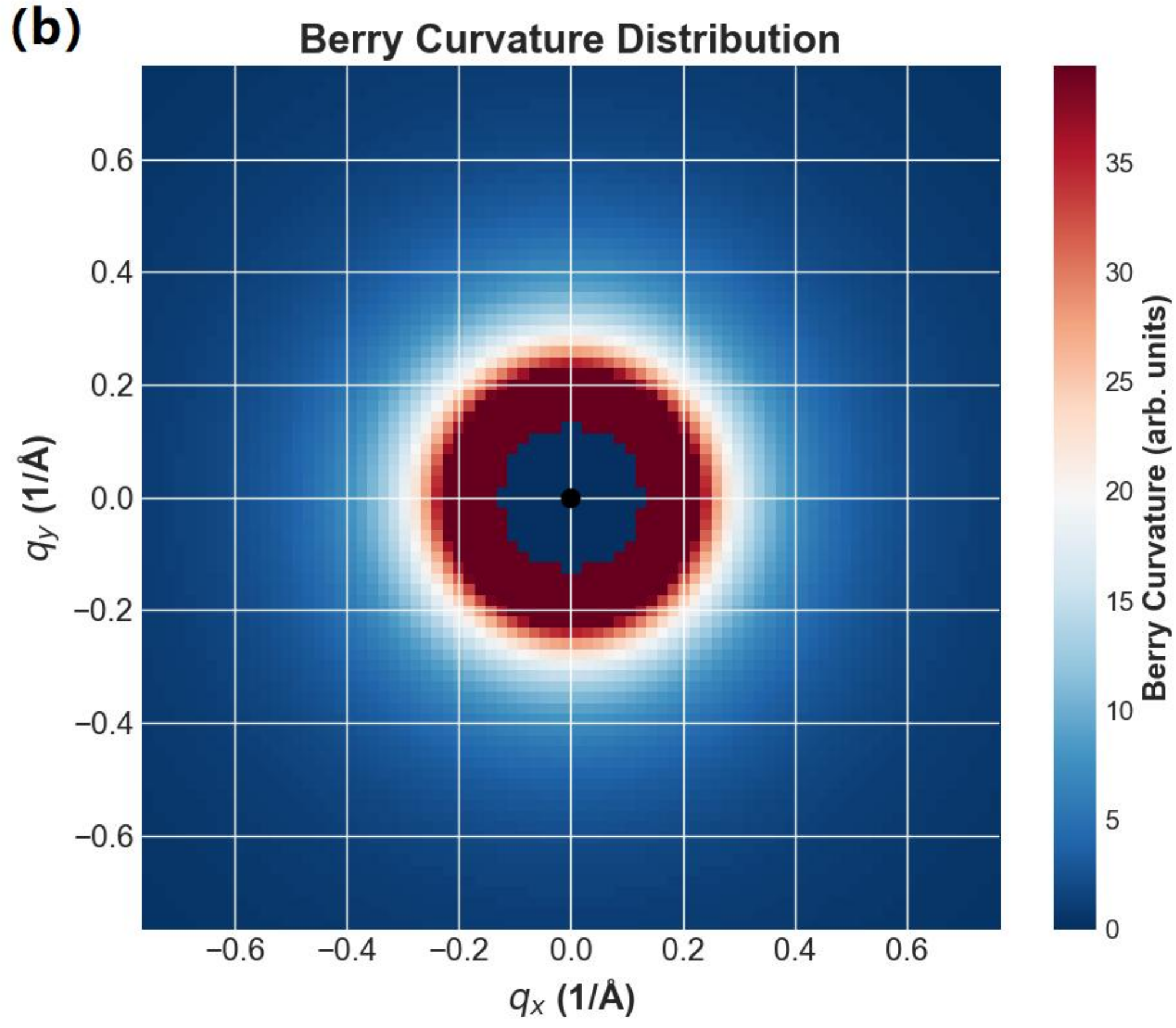

**Figure 6** (a) Phase distribution of the conduction band electron wave function on the B lattice component; (b) Distribution of Berry curvature in momentum space

**Figure 5(a)** shows the phase distribution of the conduction band electron wave function on the B lattice component. The color change represents the phase value (-π to π), and it is clear that the phase continuously changes around the Dirac point, completing one 2π orbit. This phase vortex has an integer winding number (winding number+1), which is direct evidence of the non-trivial topological structure of the wave function. The phase singularity is located at the Dirac point, where the corresponding wave function is undefined, resulting in the particularity of the zero-energy state. This vortex structure is the geometric origin of the π Berry phase.

**Figure 5(b)** shows the distribution of Berry curvature in momentum space. The Berry curvature exhibits a sharp peak near the Dirac point and rapidly decays with increasing distance. Its sign is opposite at points $K$ and $K'$, reflecting the valley degrees of freedom brought about by time reversal symmetry. This curvature distribution indicates that electrons undergo a non-trivial geometric phase accumulation in momentum space. Following a closed path of adiabatic motion around the Dirac point for one revolution results in a Berry phase of π. This topological phase directly leads to the appearance of a semi-integer quantization plateau ($\sigma_{xy}$ = ± $4e^2/h(n+1/2)$) in graphene quantum Hall conductance, which is different from traditional two-dimensional electron gases.

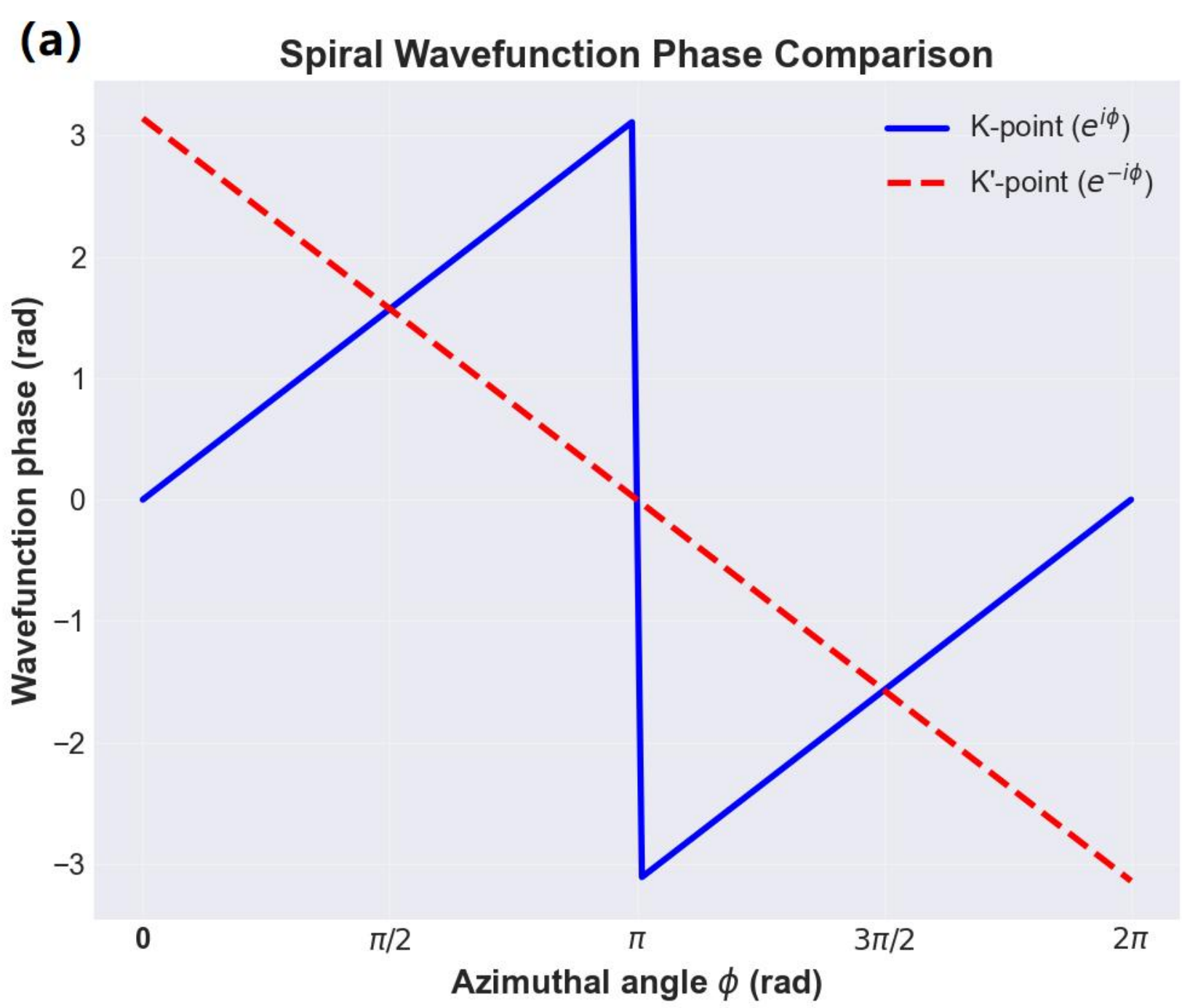

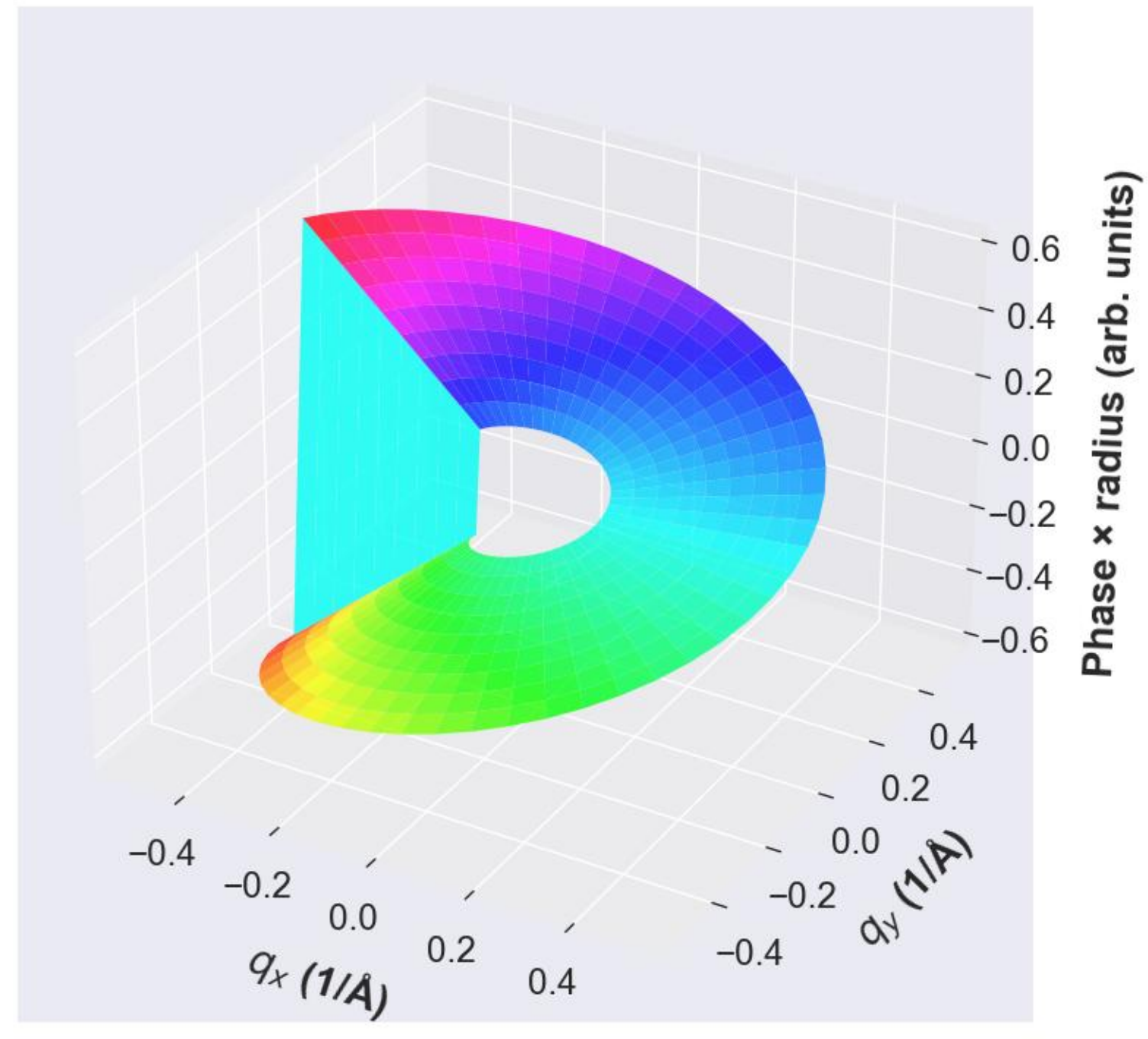

**Figure 6** (a) Phase variation of wave functions at point K and K' with respect to azimuth angle; (b) 3D Spiral Phase

**Figure 6(a)** compares the phase variation of the wave functions at point K and point K 'with respect to the azimuth angle. The phase of the conduction band wave function at point K increases linearly with ϕ ( $e^{i\phi_q}$ ), while at point K 'it decreases linearly ( $e^{-i\phi_q}$ ), indicating that the two valleys have opposite chirality. This valley-dependent spiral structure leads to pseudo-spin and momentum-direction locking of electrons during transport, strongly suppressing backscattering and playing an important role in the high mobility of graphene. The 3D spiral phase visualization in Figure 8b further highlights the continuous spiral structure of the phase in momentum space.

The low-energy effective wave function of graphene can be expressed as $\Psi(q)=1/\sqrt{2}\left(1,e^{i\phi_q}\right)^T$ , where the phase factor $e^{i\phi_q}$ reflects the helical nature of the wave function. This mathematical form directly leads to: 1) pseudo spin and momentum locking; 2) Momentum space phase vortex; 3) π Berry phase; 4) Valley degree of freedom. These topological characteristics collectively contribute to the unique transport behavior of graphene, such as weak anti-localization loss, minimum conductivity plateau, and room temperature quantum Hall effect. The concept of spiral wave function is not only applicable to graphene, but also provides a universal framework for understanding other Dirac materials, such as topological insulating surface states.

In summary, we demonstrated through systematic deduction that the low-energy quasiparticle state of graphene can be described by a wave function with spiral vortex characteristics. This wave function originates from the symmetry of the graphene honeycomb lattice, leading to a series of topological properties such as pseudo spin momentum locking, π Berry phase, and valley chirality. Together, these properties constitute the microscopic mechanism behind graphene's peculiar electron transport behavior. This study solidifies our understanding of graphene as a prototype Dirac material, and its theoretical paradigm—wave functions with vortex textures in momentum space and their related topological invariants—has been extended to the field of topological materials. For example, similar wave functions can be observed near the intersection of the surface states of topological insulators and the energy bands of topological semimetals[19]. **(See Supplemental Material)**

## 4. Conclusion

This study employs the BBC and an improved BBC model to systematically investigate graphene's electronic structure, elucidating the electrostatic screening effect, evolution of band structure and density of states (DOS), and topologically non-trivial properties. The key conclusions are:

1 .The screening parameter $\sigma_v$ quantitatively tunes electrostatic interactions. The BBC potential effectively suppresses electron-electron interactions. In the improved model, the hopping integral decays with distance and is sensitive to $\sigma_v$, while the on-site energy increases weakly with $\sigma_v$. Both models eliminate the Coulomb singularity, with hopping integrals vanishing at long range.

2.$\sigma_v$ directs band renormalization: energies at high-symmetry points decrease with $\sigma_v$ in the BBC model, while showing opposite trends at Γ and K in the improved model, indicating momentum selectivity. Band narrowing occurs, and a bandgap may open for $\sigma_v \geq 1.00$. The DOS peaks at the Fermi level, with the high-energy region slightly increasing with $\sigma_v$.

3.The low-energy states of graphene are described by helical wavefunctions, yielding topological properties like pseudospin-momentum locking and a π Berry phase, which explain unique transport phenomena.

## Acknowledgements

This study was financially supported by four research programs: 1) the Science and Technology Research Project of Fuling District, Chongqing Municipality (Grant No. FLKJ-2024BAG5128); 2) the Science and Technology Research Program of Chongqing Municipal Education Commission (Grant No. KJZD-K202501406); 3) Program of the Chongqing Municipal Education Commission (Grant No.KJQN-202401420) ;4) the Science and Technology Research Project of Fuling District, Chongqing Municipality (Grant No. FLKJ-2024BAG5133).

# Supplemental Material

## Electrostatic Screening Modulation of Graphene's Electronic Structure and the Helical Wavefunction-Dominated Topological Properties

Yaorui Tan ,Xiang Chen, Yunhu Zhu, Xiaowu Yang, Zhongkai Huang, Chuang Yao, Maolin Bo*

*Key Laboratory of Extraordinary Bond Engineering and Advanced Materials Technology (EBEAM) of Chongqing, College of Mechanical and Electrical Engineering, Yangtze Normal University, Chongqing 408100, China*

*Corresponding author: Maolin Bo (E-mail addresses: bmlwd@yznu.edu.cn)

## 1. Quantum many-body systems

A many-body system composed of atomic nuclei and electrons. Its non-relativistic Schrödinger equation is:

$$\hat{H}\Psi\ (\{r_i\},\{R_A\}) = E\Psi$$

The Hamiltonian $\hat{H}$ includes the kinetic energy term, nucleus nucleus repulsion, electron nucleus attraction, and electron electron Coulomb interaction

$$\hat{H} = -\sum_{i}\frac{\hbar^2}{2M_l}\nabla_i^2 - \sum_{i}\frac{\hbar^2}{2m_e}\nabla_i^2 - \sum_{i,A}\frac{Z_A e^2}{4\pi\ _0\left|r_i - R_A\right|} + \frac{1}{2}\sum_{i\neq j}\frac{e^2}{4\pi\epsilon_0\left|r_i - r_j\right|} + \frac{1}{2}\sum_{A\neq B}\frac{Z_A Z_B e^2}{4\pi\ _0\left|R_A - R_B\right|}$$

To simplify the problem, we introduce the single electron approximation, which assumes that each electron moves independently in an average effective potential field $V_{eff}(r)$ generated jointly by the atomic nucleus and other electrons. The wave equation for a single electron is:

$$\left[-\frac{\hbar^2}{2m_e}\nabla^2 + V_{eff}(r)\right]\psi_i(r) = \epsilon_i\psi_i(r)$$

The total electron energy $E_{el}$ of the system is obtained by summing the single electron energy levels (eigenvalues) of the occupied states

$$E_{el} = \sum_{i}^{occ}\epsilon_i$$

Note that $V_{eff}$ here is a functional of electron density $\rho$(r) and needs to be iteratively solved through self consistent field (SCF), which is the source of expensive density functional theory (DFT) calculations.

To obtain analytical or semi analytical models, we further introduce linear combination basis sets of atomic orbitals and tight binding approximations. We will expand the single electron wave function using a set of localized atomic orbital basis functions $\{\phi_\mu(r)\}$:

$$\psi_i(r) = \sum_\mu c_{\mu i}\phi_\mu(r)$$

Among them, each atomic orbital $\phi_\mu$ is tightly localized near a certain atomic nucleus *A(μ)*. Substitute the expansion equation into the single electron equation and multiply it by $\phi_v^*$ for integration to obtain the generalized eigenvalue problem:

$$\sum_\mu (H_{v\mu} - \epsilon_i S_{v\mu}) c_{\mu i} = 0$$

Here, $S_{v\mu} = \langle \phi_v | \phi_\mu \rangle$ is the overlapping matrix, while Hamiltonian matrix element $H_{v\mu}$ is the key:

$$H_{v\mu} = \left\langle \phi_v | \hat{T} + V_{eff} | \phi_\mu \right\rangle = T_{v\mu} + V_{v\mu}$$

Double center integral approximation: The values of $\mathrm{H}_{v\mu}$ and $S_{v\mu}$ depend only on the properties and relative positions of the two atoms to which orbitals $v$ and $\mu$ belong, and are independent of other atoms in the system. Short range truncation: When the distance between two atoms exceeds a certain truncation radius, the matrix element becomes zero.

**2. Traditional BBC Model - Specific Parameterization of Effective Potential**

The original BBC model is an empirical and parameterized modeling of the effective potential matrix element $V_{v\mu} = \langle \phi_v | V_{eff} | \phi_\mu \rangle$ in the tight constraint framework mentioned above. The diagonal element $\mathrm{H}_{\mu\mu} = \varepsilon_\mu$ is called the atomic orbital energy level. The BBC model models it as:

$$\varepsilon_\mu \approx \left\langle \phi_\mu \,|\, \hat{T} + V_A^{BBC} \,|\, \phi_\mu \right\rangle$$

Among them, $V_A^{BBC}$ is a pre-set mathematical form of the shielding Coulomb potential (Yukawa potential) centered on atom $A(\mu)$:

$$V_A^{BBC}(r) = -\frac{Z_A - \mu_v}{|r - R_A|} e^{-\mu_0 |r - R_A| / a_0}$$

Physical interpretation: This potential attempts to simulate the average potential felt by an electron near atom A using two terms: $-\frac{Z_A - \mu_v}{|r - R_A|}$ : simulates the attraction of a partially shielded atomic nucleus. $\mu_v$ is the shielding constant, representing the amount of nuclear charge shielded by the inner and valence electrons. $e^{-\mu_0 |r - R_A| / a_0}$ : The Yukawa shielding factor simulates the presence of other electrons, causing the potential field to decay exponentially with distance, rather than extending over long distances like the bare core potential. This is equivalent to packaging the shielding effects of other electronic clouds into this attenuation factor.

The non diagonal element $\mathrm{H}_{v\mu}(v \neq \mu)$ describes the "hopping" of electrons between different atomic orbitals and is the source of chemical bond formation. In BBC like models, it is typically parameterized as a function of diagonal terms and overlapping integrals, such as the extended Hückel form approximation:

$$\mathrm{H}_{v\mu} \approx \frac{1}{2} K (\varepsilon_v + \varepsilon_\mu) S_{v\mu}$$

Among them, *K* is an empirical constant. After obtaining the Hamiltonian matrix and diagonalizing it, the single electron energy level { $\epsilon_i$ } is obtained, with a total electron energy of:

$$E_{el}^{BBC} = \sum_{i}^{occ} \epsilon_i$$

The total energy needs to be added to the nuclear nuclear repulsion energy $E_{nn}$, which is usually calculated using the shielded Coulomb form:

$$E_{tot}^{BBC} = E_{el}^{BBC} + \frac{1}{2}\sum_{A \neq B} \frac{Z_A Z_B - \mu_{nn}^2}{|R_A - R_B|} e^{-\mu_{nn}|R_A - R_B/a_0|}$$

The core limitation of traditional BBC is that the shielding constants $\mu_v$ and $\mu_{nn}$ are global constants for each element. This means that the potential function form of a carbon atom is completely the same whether it is in diamond, graphene, benzene molecules, or nanotubes. This completely ignores the significant impact of chemical environment on electronic structure and shielding effect, which is the main reason for its poor accuracy and transferability.

**3. ML-BBC Model - Machine Learning Introducing Environment Dependent Intelligent Shielding Parameters**

The ML-BBC model introduces environment dependent intelligent masking parameters through machine learning, fundamentally correcting the aforementioned shortcomings of traditional BBC.

The core idea of ML-BBC is that the shielding constant $\mu_v^A$ should not be a global constant, but a function that depends on the local chemical environment of atom A.

$$\mu_v^A = \mathcal{F}_{ML}(\mathcal{D}_A)$$

Among them, $\mathcal{D}_A$ is a mathematical descriptor that describes the local environment of atom A, and $\mathcal{F}_{ML}$ is a machine learning model (such as a neural network). To ensure the rotation, translation, and permutation symmetry of the model, we use SOAP descriptors. Its construction is a rigorous multi-step mathematical process:

(1) Neighborhood atomic density representation: Truncate the neighbors within the radius of atom A as a continuous, element resolved atomic density function:

$$\rho_A(r) = \sum_{i \in N(A)} \delta(Z_i) \exp\left( -\frac{|r - r_i|^2}{2\sigma^2} \right) f_{cut}(r_{iA})$$

(2) Spherical harmonic and radial basis expansion: Expand the density function in a spherical coordinate system with A as the origin, using spherical harmonic function $Y_{lm}(\theta,\phi)$ and a set of orthogonal radial basis functions $g_n(r)$:

$$\rho_A^H(r,\theta,\phi) \approx \sum_{n=0}^{n_{\max}} \sum_{l=0}^{l_{\max}} \sum_{m=-l}^{l} c_{nlm}^Z \cdot g_n(r) \cdot Y_l^m(\theta,\phi)$$

The coefficient $c_{nlm}$ is obtained by the integral projection of the density function and the basis function.

(3) Constructing rotation invariant: By calculating the power spectra of all combinations of n, n', l, obtain descriptor components that are invariant to rotation:

$$p_{nn'l}^{ZZ'} = \pi \sqrt{\frac{8}{2l+1}} \sum_{m=-l}^{l} \left( c_{nlm}^Z \right)^* c_{n'lm}^{Z'}$$

(4) Multi element extension: For systems containing multiple elements, establish independent density channels $\rho_A^Z$ for each element Z, and calculate the power spectrum $p_{nn'l}^{ZZ'}$ across element channels. The final descriptor vector $\mathcal{D}_A$ is the concatenation of all $p_{nn'l}^{ZZ'}$. The final descriptor vector $\mathcal{D}_A$ is the scalar $p_{nn'l}^{ZZ'}$ corresponding to all meaningful exponential combinations (Z, Z', n, n', l), concatenated in a certain order to form a one-dimensional long vector:

$$\mathcal{D}_A = \left[ \ldots, p_{nn'l}^{ZZ'}, \ldots \right]^T$$

We use a multi-layer perceptron (MLP) or graph neural network (GNN) to learn the mapping from high-dimensional descriptor $\mathcal{D}_A$ to scalar masking parameter $\mu_v^A$. This network is applied to every atom in the system in a weight sharing manner.

Input layer： $\mathcal{D}_A$

Hidden layer： $h_A^{(k)} = \mathrm{Re}LU(W^k h_A^{(k-1)} + b^{(k)})$

Output layer： $\mu_v^A = \sigma(W^{(out)} h_A^{final} + b^{(out)})$, where σ is a Sigmoid function, limiting the output to a reasonable range (such as [0.5, 5.0]).

In order to achieve high computational efficiency and facilitate integration with DFT data, ML-BBC adopts the energy calculation paradigm of modern machine learning potentials (MLIP), abandoning the steps of explicit construction and diagonalization of Hamiltonian matrices. ML-BBC makes a key assumption: the total energy of the system can be decomposed into the sum of the "atomic energy" $E_A$ contributed by each atom:

$$E_{tot} = \sum_{A=1}^{N} E_A$$

This is a proposed approach, but it has been proven to be extremely effective in describing the potential energy surfaces of many systems. The atomic energy $E_A$ is determined by its local environment. We use the second small neural network $MLP_E$, with the masking parameter $\mu_v^A$ and the hidden layer feature $h_A$ of the main regression network as inputs, to directly predict $E_A$:

$$E_A = MLP_E\left(\mu_v^A, h_A, OneHot(Z_A)\right)$$

Among them, $OneHot(Z_A)$ is a unique encoding of atomic species, providing element identity information for the network. Atomic force is the negative gradient of total energy with respect to atomic position. Thanks to the full process automatic differentiation of descriptors, neural networks, and energy summation, we can accurately and efficiently calculate:

$$F_i = -\frac{\partial E_{tot}}{\partial R_i} = -\sum_A \frac{\partial E_A}{\partial R_i}$$

The calculation of force does not rely on any finite difference and is analytical, which is crucial for conducting stable molecular dynamics simulations.

### 4. Loss function

The model parameters are learned by minimizing the following loss function:

$$\mathcal{L}(\Theta)=\frac{1}{N_{\text{batch}}}\sum_{k=1}^{N_{\text{batch}}}\left[\frac{\lambda_E}{N_{atom}^{(k)}}\left(E_k^{DFT}-E_k^{ML-BBC}(\Theta)^2\right)+\frac{\lambda_E}{3N_{atom}^{(k)}}\sum_{i=1}^{N_{\text{atom}}^{(k)}}\left\|\mathbf{F}_{i,k}^{DFT}-\mathbf{F}_{i,k}^{ML-BBC}(\Theta)\right\|^2+\lambda_S\left\|\sigma_k^{DFT}-\sigma_k^{ML-BBC}(\Theta)\right\|_F^2\right]+\lambda_{reg}R(\Theta)$$

Detailed explanation of formula variables

| variable | Meaning and Explanation |
|---|---|
| $\Theta$ | The set of all trainable parameters in the model, including the weights and biases of neural network $\mathcal{F}_{ML}$ and $MLP_E$. |
| $N_{batch}$ | The number of samples in a batch during training. |
| $N_{atom}^{(k)}$ | The total number of atoms in the kth configuration. |
| $E_k$ | The total energy of the k-th configuration. The superscript DFT represents high-precision calculation labels, while ML-BBC represents model predicted values. |
| $\mathbf{F}_{i,k}$ | The force (three-dimensional vector) exerted on the i-th atom in the kth configuration. |
| $\sigma_k$ | The virial stress tensor of the kth configuration (usually a 3×3 matrix, crucial for periodic solid systems). |

| variable | Meaning and Explanation |
| --- | --- |
| $\left\Vert \cdot \right\Vert_F$ | The Frobenius norm, defined as $\left\Vert A \right\Vert_F = \sqrt{\sum_{i,j} A_{ij}^2}$ for matrix A, is used to measure the total error of the stress tensor. |
| $R(\Theta)$ | Regularization term, commonly in the form of L2 regularization: $R(\Theta) = \sum_{\theta \in \Theta} \theta^2$ |
| $\lambda_E, \lambda_F, \lambda_S, \lambda_{reg}$ | Superparameters are used to balance the relative weights between the four losses. |

**5. ML-BBC model electron-phonon coupling Hamiltonian**

Expansion of electronic Hamiltonian on atomic position:

$$\hat{H}_e\left(\left\{\mathbf{R}_A\right\}\right) = \hat{H}_e^0 + \sum_{A,\alpha} \frac{\partial \hat{H}_e}{\partial R_{A\alpha}} \delta R_{A\alpha} + \frac{1}{2} \sum_{A\alpha, B\beta} \frac{\partial^2 \hat{H}_e}{\partial R_{A\alpha} \partial R_{B\beta}} \delta R_{A\alpha} \delta R_{B\beta} + \cdots$$

The first-order term is the electron phonon coupling term. Atomic displacement is represented by phonon operator:

$$\delta R_{A\alpha} = \frac{1}{\sqrt{N}} \sum_{\mathbf{q},\nu} \sqrt{\frac{\hbar}{2M_A \omega_{\mathbf{q}\nu}}} e_{A\alpha}\left(\mathbf{q}\nu\right)\left(b_{\mathbf{q}\nu} + b_{-\mathbf{q}\nu}^{\dagger}\right)$$

The intrinsic problem corresponding to the dynamic matrix is:

$$\sum_{B\beta} D_{A\alpha,B\beta}\left(\mathbf{q}\right) e_{B\beta}\left(\mathbf{q}\nu\right) = \omega_{\mathbf{q}\nu}^2 e_{A\alpha}\left(\mathbf{q}\nu\right)$$

Under the tight constraints:

$$\hat{H}_e = \sum_{ij,\sigma} t_{ij} \hat{c}_{i\sigma}^{\dagger} \hat{c}_{j\sigma}$$

The phonon coupling Hamiltonian can be written as:

$$\hat{H}_{e-ph} = \frac{1}{\sqrt{N}} \sum_{\substack{ij,\sigma \\ \mathbf{q},\nu}} g_{ij}^{\mathbf{q}\nu} \hat{c}_{i\sigma}^{\dagger} \hat{c}_{j\sigma} \left( \hat{b}_{\mathbf{q}\nu} + \hat{b}_{-\mathbf{q}\nu}^{\dagger} \right)$$

The expression for its coupling constant is:

$$g_{ij}^{\mathbf{q}\nu} = \sum_{A,\alpha} \frac{\partial t_{ij}}{\partial R_{A\alpha}} \sqrt{\frac{\hbar}{2M_A \omega_{\mathbf{q}\nu}}} e_{A\alpha}(\mathbf{q}\nu)$$

**5.1 ML-BBC model jump integral predictor**

The basic ML-BBC model can perceive the environment and predict shielding parameters through functions $\mu_\nu^A = F_{ML}(D_A)$, and then calculate atomic energy $E_A$ and total energy $E_{tot}$. To calculate electron phonon coupling, we need to understand how electrons "sense" the small displacement of atoms, that is, how the electron Hamiltonian $H$ varies with the atomic position $R_A$. Under the tight constraint framework, this boils down to calculating the derivatives of diagonal elements (orbital energy levels $\varepsilon_i$) and non diagonal elements (jump integrals $t_{ij}$)with respect to $R_A$.

We introduce a specially designed and differentiable neural network module - the jump integral predictor $\mathcal{F}_{\text{hop}}$: $\mathcal{F}_{\text{hop}}(\mathcal{D}_i, \mathcal{D}_j, \mathbf{R}_{ij}; \Phi)$ is strictly defined physically: the jump integral $t_{ij}$ is uniquely determined by the complete local chemical environment ($D_i$, $D_j$) of two atoms and their relative positions ($R_{ij}$). $t_{ij} = \mathcal{F}_{\text{hop}}(\mathbf{h}_i, \mathbf{h}_j, \mathcal{G}(\mathbf{R}_{ij}); \Phi)$ is an optimized engineering implementation that specifically explains how to efficiently and accurately implement the above physical definitions.

Input instructions:

1. $\mathbf{h}_i$, $\mathbf{h}_j$: They are the environmental eigenvectors of atoms $i$ and $j$. This is not the original SOAP descriptor $D$, but the output of the intermediate or final hidden layer of the main regression network $F_{ML}$. $\mathbf{h}$ contains richer high-order environmental information abstracted by neural networks than $\mu_v$. Here, $h_i$ is used instead of $D_i$: in fact, $h_i = F_{ML}(D_i)$ is the high-dimensional feature representation of $D_i$ extracted by the main network, which has lower dimensions but contains core information related to energy, force, and chemical bonds.

2. $\mathcal{G}(\mathbf{R}_{ij})$ ： Geometric encoding of vector $\mathbf{R}_{ij} = R_j - R_i$ between atoms. Directly using (x, y, z) coordinates will break rotational symmetry, so radial basis function sets are usually used for encoding:

$$\mathcal{G}(\mathbf{R})_k = \exp\left(-\gamma_k(\| \mathbf{R}_{ij} \| - r_k)^2\right) \cdot f_{\text{cut}}(\| \mathbf{R}_{ij} \|)$$

Among them, $(\gamma_k - r_k)$ is the preset parameter set. In this way, $\mathcal{G}(\mathbf{R})$ is a vector that is only related to distance, ensuring rotational invariance. $\mathcal{G}(\mathbf{R}_{ij})$ replaces $R_{ij}$: $\mathcal{G}(\mathbf{R}_{ij})$ is an encoding function that converts the vector $R_{ij}$ into a set of rotation invariant features, ensuring the physical symmetry of the model. The original coordinates $R_{ij}$ themselves do not satisfy rotational invariance.

Network Design:

$\mathcal{F}_{\text{hop}}$ is usually a multi-layer perceptron (MLP). The input layer concatenates the vectors $\mathbf{h}_i$, $\mathbf{h}_j$, and $\mathcal{G}(\mathbf{R}_{ij})$. This "dual stream" input design allows the network to simultaneously consider the complex chemical environment of two sites and their spatial geometric relationships.

5.1.2 Training Strategy and Joint Optimization

Training this extended model requires multi task learning.

1. New training data: In addition to energy E and force F, it is now necessary to prepare a reference jump integral $t_{ij}^{ref}$ as a label for each training configuration. These labels are typically obtained through DFT calculations combined with Wannier90's Maximum Localized Wannier Function (MLWF) method, which can fit the optimal tight binding parameters from first principles wave functions.

2. Extended loss function:

$$\mathcal{L}_{\text{total}} = \lambda_E \mathcal{L}_E + \lambda_F \mathcal{L}_F + \lambda_t \mathcal{L}_t + \lambda_{\text{reg}} \mathcal{L}_{\text{reg}}$$

$\mathcal{P}_k$ is the set of atomic pairs to be considered in the *k*-th configuration (usually defined by truncation radius). $w_{ij}$ is an optional weight that can be used to emphasize the importance of nearest neighbor keys or specific types of keys. The set of all trainable parameters for the model $\Theta$. The specific forms of each item are:

$$\mathcal{L}_E = \frac{1}{N_{\text{batch}}} \sum_{k=1}^{N_{\text{batch}}} \left( E^{ref}(K) - E^{\text{ML-BBC}}(k) \right)^2$$

$$\mathcal{L}_F = \frac{1}{N_{\text{batch}} N_{\text{atom}}} \sum_{k=1}^{N_{\text{batch}}} \sum_{i=1}^{N_{\text{atom}}} \left\| \mathbf{F}_i^{ref}(k) - \mathbf{F}_i^{\text{ML-BBC}}(k) \right\|^2$$

$$\mathcal{L}_t = \frac{1}{N_{\text{batch}} N_{\text{pair}}} \sum_{k=1}^{N_{\text{batch}}} \sum_{\langle ij \rangle \in \mathcal{P}_k} w_{ij} \left( t_{ij}^{\text{ref}}(k) - t_{ij}^{\text{ML-BBC}}(k) \right)^2$$

$$\mathcal{L}_{\text{reg}} = \sum_{\theta \in \Theta} |\Theta|^2 \ (L2) or \ \mathcal{L}_{\text{reg}} = \sum_{\theta \in \Theta} |\Theta| \ (L1)$$

3. Training process: All parameters ($\Theta$ of main network $F_{ML}$, parameters of energy network $MLP_E$, and $\Phi$ of jump integral predictor $\mathcal{F}_{\text{hop}}$) are jointly optimized simultaneously. This forces the model to learn features $h_i$ and $\mu_v^i$ that are simultaneously effective for predicting total energy, atomic forces, and electron jump integrals, ensuring physical consistency between different outputs.

### 5.2 Construction of Electronic Hamiltonian and Calculation of Electronic States

The extended ML BBC model can predict the electronic structure of a given atomic configuration $R_A$ in real-time.

1. Construct a tightly bound Hamiltonian matrix $H$ :

Diagonal element: $H_{ii} = \varepsilon_i$. The orbital energy level $\varepsilon_i$ can be directly derived from shielding parameters or mapped through another small network to obtain: $\varepsilon_i = f\left(\mu_v^i, h_i, Z_i\right)$.

Non diagonal elements: for $i \neq j$, $H_{ii} = t_{ij}$, directly predicted by $\mathcal{F}_{\text{hop}}$. The matrix $H$ is Hermitian $\left(H_{ij} = H_{ji}^*\right)$, which requires the design of $\mathcal{F}_{\text{hop}}$ to ensure output symmetry.

2. Diagonalization to obtain electronic states:

$$H\mathbf{c}_n = \epsilon_n \mathbf{c}_n$$

By diagonalizing the matrix of $N_{basis} \times N_{basis}$ (where $N_{basis}$ is the number of basis functions), we obtain the electron eigenspectrum $\varepsilon_n$. Eigenvector $\mathbf{c}_n$, where $\mathbf{c}_i^n$ is the expansion coefficient of the *n*-th eigenstate on the *i*-th atomic orbital basis. Fermi level $E_F$ : determined by filling the total number of electrons: $N_{\mathrm{e}} = 2\sum_n f\left(\frac{\varepsilon_n - E_F}{k_B}T\right)$, where $f$ is the Fermi Dirac distribution function.

### 5.3 Calculation of electro acoustic coupling matrix element based on automatic differentiation

This is a key innovative step in combining machine learning potential with quantum perturbation theory, with the goal of calculating the electron phonon coupling matrix elements:

$$g_{mn}^{\mathbf{q}\nu} = \langle \psi_m | \frac{\partial V}{\partial u_{\mathbf{q}\nu}} | \psi_n \rangle$$

Among them, $\frac{\partial V}{\partial u_{\mathbf{q}\nu}}$ is the derivative of the phonon mode displacement of the crystal potential along the wave vector $\mathbf{q}$ and branch $\nu$.

We adopt a hybrid method of "finite displacement+automatic differentiation", which has a precise and efficient process:

Step 1: Calculate the phonon spectrum (Harmonic Phonons)

Firstly, it is necessary to know the lattice vibration mode of the system.

1. Calculate the force constant matrix: Utilize the accurate force prediction ability of the basic ML-BBC.

$$\Phi_{A\alpha,B\beta} = -\frac{\partial F_{A\alpha}}{\partial R_{B\beta}} = \frac{\partial^2 E_{\text{tot}}}{\partial R_{A\alpha} \partial R_{B\beta}}$$

The second derivative of energy can be directly calculated through automatic differentiation, or the finite displacement method can be used (applying a small displacement $\pm\delta$ to atom B and calculating the change in force on atom A using ML-BBC).

2. Construct and diagonalize the dynamic matrix:

$$D_{A\alpha,B\beta}(\mathbf{q}) = \frac{1}{\sqrt{M_A M_B}} \sum_{\mathbf{R}} \Phi_{0\alpha,\mathbf{R}\beta} e^{i\mathbf{q}\cdot\mathbf{R}}$$

$$D(\mathbf{q})\mathbf{e}(\mathbf{q}\nu) = \omega_{\mathbf{q}\nu}^2 \mathbf{e}(\mathbf{q}\nu)$$

Output: phonon frequency $\omega_{\mathbf{q}\nu}$ and polarization vector $\mathbf{e}_{A\alpha}(\mathbf{q}\nu)$.

Step 2: Generate perturbation configurations along phonon modes

For the phonon mode ($\boldsymbol{q},\boldsymbol{v}$) that needs to be calculated, generate a set of supercell configurations distorted according to the shape of that mode:

$$\mathbf{R}_A(\lambda) = \mathbf{R}_A^0 + \lambda \cdot \mathbf{e}_A(\mathbf{q}\nu)$$

Among them, $\lambda$ is the mode amplitude (take the small value, such as ± 0.01 Å). Usually, three points $\lambda = 0, \pm\delta$ can be calculated.

Step 3: Use automatic differentiation to calculate $\dfrac{\partial H_{ij}}{\partial \lambda}$

This is the most crucial step, utilizing the complete differentiability of the extended ML BBC model.

1. For the configuration of $\lambda = \pm\delta$, quickly calculate its tightly bound Hamiltonian matrix using a model $H(\lambda)$。

2. In the process of calculating $H(\lambda)$, the entire calculation graph (from atomic coordinates $R_A(\lambda)$, to descriptors, to network transmission, to outputs $t_{ij}$ and $\varepsilon_i$, and finally assembled into $H$) is automatically recorded.

3. Using automatic differentiation (backpropagation), directly calculate the gradient of amplitude $\lambda$ for each matrix element $H_{ij}$ in $H$:

$$\frac{\partial H_{ij}}{\partial \lambda}\Big|_{\lambda=0} = \text{autograd}\left(H_{ij}, \lambda\right)$$

Advantages: Compared to traditional finite difference methods $\dfrac{H(+\delta) - H(-\delta)}{2\delta}$, the gradient obtained by automatic differentiation is analytical and more accurate, and all gradients of $H_{ij}$ to $\lambda$ can be obtained by computing the backpropagation once, with extremely high efficiency.

Step 4: Basic transformation yields $g_{mn}^{\mathbf{q}\nu}$

Diagonalization at equilibrium position $(\lambda = 0)$ yields the electronic eigenstates $\psi_m$ and coefficients $c_i^m$. Transform the derivative $\frac{\partial H_{ij}}{\partial \lambda}$ calculated in step 3 from the atomic orbital basis to the eigenstate basis:

$$g_{mn}^{\mathbf{q}\nu} = \sum_{i,j} (c_i^m)^* \left( \frac{\partial H_{ij}}{\partial \lambda} |_{\lambda=0} \right) c_j^n$$

$g_{mn}^{\mathbf{q}\nu}$ is the electron phonon coupling matrix element that connects electronic states m and n with phonon mode $(\mathbf{q}, \nu)$. It contains the normalization factor of phonon modes, whose dimension is energy.

**5.4 Calculation and physical significance of macroscopic physical quantities**

After obtaining the micro coupling matrix elements, macroscopic quantities with direct physical significance can be obtained through integration.

**5.4.1 Eliashberg spectral function** $\alpha^2 F(\omega)$

This is a key function that describes the contribution weights of phonons of different frequencies to the coupling of electroacoustic particles.

$$\alpha^2 F(\omega) = \frac{1}{N(E_F)} \sum_{\mathbf{q},\nu} \frac{\omega_{\mathbf{q}\nu}}{2\pi N_q} \sum_{mn} |g_{mn}^{\mathbf{q}\nu}|^2 \ \delta(\ _m - E_F)\delta(\ _n - E_F)\delta(\omega - \omega_{\mathbf{q}\nu})$$

$N(E_F)$ : The density of electronic states at the Fermi level. $N_q$ : Number of sampling $q$ points. Physical meaning: The value of $\alpha^2 F(\omega)$ at frequency $\omega$ is proportional to the sum of the coupling strength between phonons and Fermi surface electrons at frequencies around $\omega$.

**5.4.2 Total electron phonon coupling constant** $\lambda$

$$\lambda = 2\int_0^\infty \frac{\alpha^2 F(\omega)}{\omega} d\omega = \frac{1}{N_q} \sum_{\mathbf{q},\nu} \lambda_{\mathbf{q}\nu}$$

The mode coupling strength is:

$$\lambda_{\mathbf{q}\nu} = \frac{1}{N(E_F)\omega_{\mathbf{q}\nu}^2} \sum_{mn} |g_{mn}^{\mathbf{q}\nu}|^2 \,\delta(\;_{m} - E_F)\delta(\;_{n} - E_F)$$

Physical meaning: $\lambda$ is a dimensionless number that measures the overall strength of electron phonon interactions. λ<0.5 is usually weak coupling, 0.5<λ<1.5 is moderate to strong coupling, and λ>1.5 is very strong coupling.

**5.4.3 Estimation of superconducting transition temperature (for electroacoustic superconductors)**

Using the classic McMillan Daines formula:

$$T_c = \frac{\omega_{\log}}{1.2} \exp\left[ -\frac{1.04(1+\lambda)}{\lambda - \mu^*(1+0.62\lambda)} \right]$$

Among them: $\omega_{\log}$ is the logarithmic mean phonon frequency:

$$\omega_{\log} = \exp\left[ \frac{2}{\lambda} \int_0^{\infty} \frac{d\omega}{\omega} \alpha^2 F(\omega) \ln \omega \right]$$

$\mu^*$ is the effective Coulomb pseudopotential, typically between 0.1 and 0.15, which is a parameter that requires empirical input or determination through higher-level calculations. Physical significance: This formula establishes a semi quantitative relationship between the phonon property $\left(\alpha^2 F(\omega), \lambda\right)$, electronic property $\left(N\left(E_F\right)\right)$, and superconducting critical temperature $T_c$ of materials.

**5.4.4 Electron scattering rate and resistivity**

Under the relaxation time approximation, the electron scattering rate determined by electron phonon coupling (reciprocal of lifetime $\tau$) is:

$$\frac{1}{\tau_{n\mathbf{k}}} = \frac{2\pi}{\hbar} \sum_{m,\mathbf{q},\nu} |g_{mn}^{\mathbf{q}\nu}|^2 \left[ (1 - f_m + n_\nu)\delta(\;_{n\mathbf{k}} - \;_{m\mathbf{k}+\mathbf{q}} + \hbar\omega_{\mathbf{q}\nu}) + (f_m + n_\nu)\delta(\;_{n\mathbf{k}} - \;_{m\mathbf{k}+\mathbf{q}} - \hbar\omega_{\mathbf{q}\nu}) \right]$$

Among them, B and C are the Fermi Dirac and Bose Einstein distribution functions, respectively. From this, the DC resistivity $\rho \propto 1/\tau$ can be calculated.

## 6. Spiral wave function and Berry curvature

In the conventional tight-binding model, the electronic dispersion near the Fermi level is generally not linear, and the underlying physical mechanism is often implicit in numerical results Fig.S1(a). Meanwhile, topological properties can only be revealed through additional calculations, lacking intuitive analytical insight.The helical wave model features prominent advantages in ultrahigh computational efficiency and a clear physical picture. This model yields analytical solutions near the Dirac point, from which topological properties such as Berry curvature and helical phase can be directly derived Fig.S1(b)(c)(d) . It also explicitly exhibits valley degrees of freedom and chiral symmetry, greatly simplifying theoretical analysis of low-energy physics.Nevertheless, its applicability is strictly limited: it is valid only in the small-momentum region (roughly $|q| < 0.2 \times 2\pi/a$), neglecting the nonlinear bending of energy bands and electron – electron interactions. Consequently, it cannot describe high-energy excitations, defect effects, or complex material modifications.

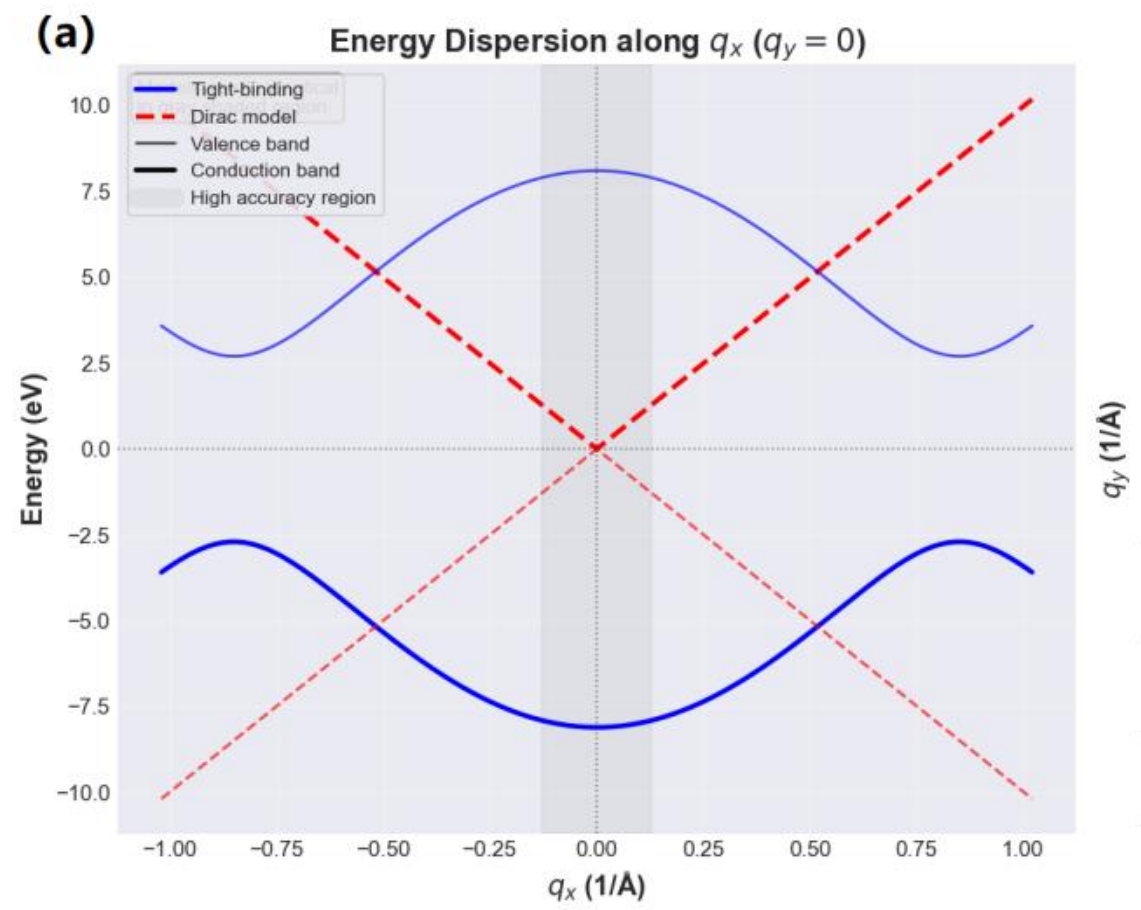

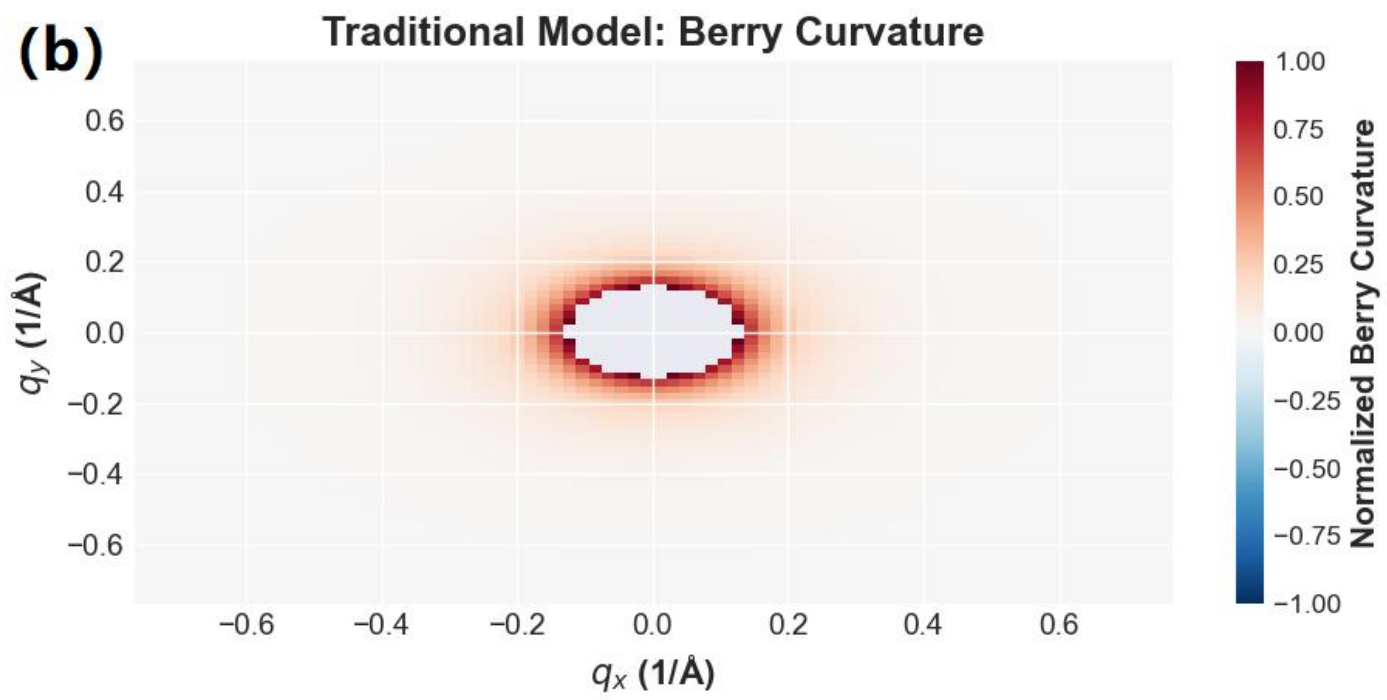

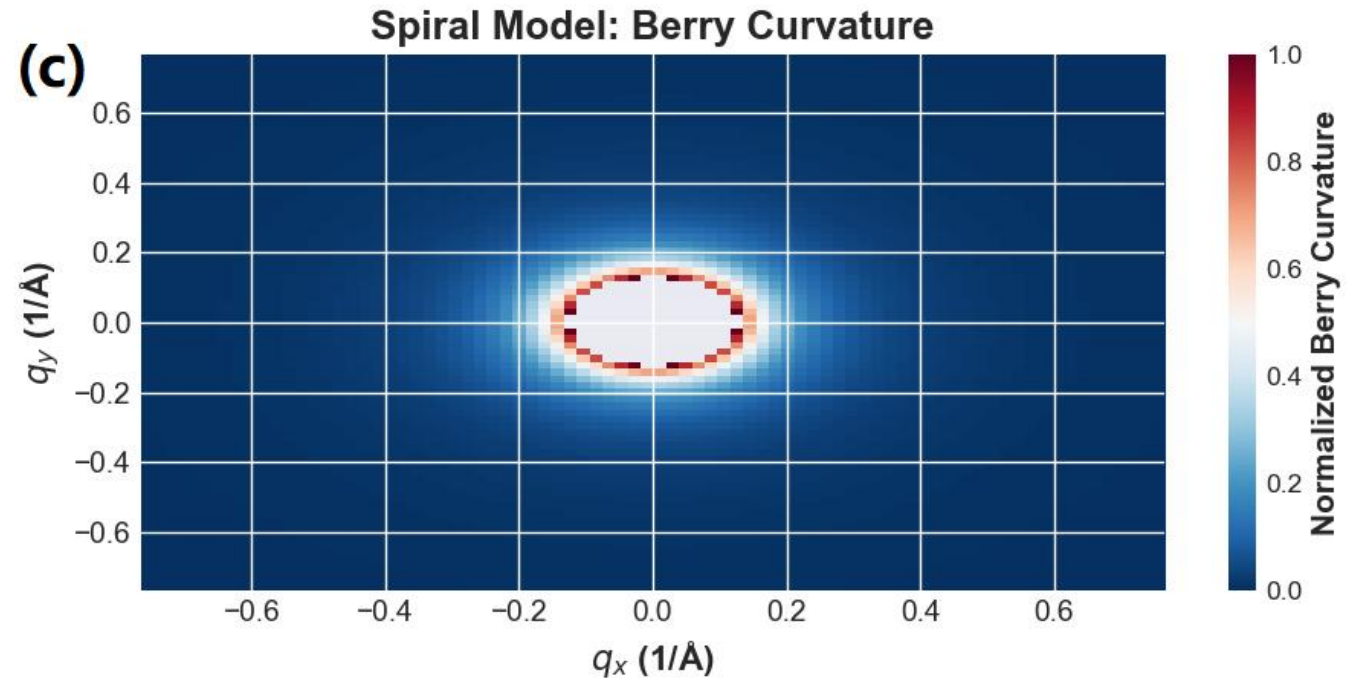

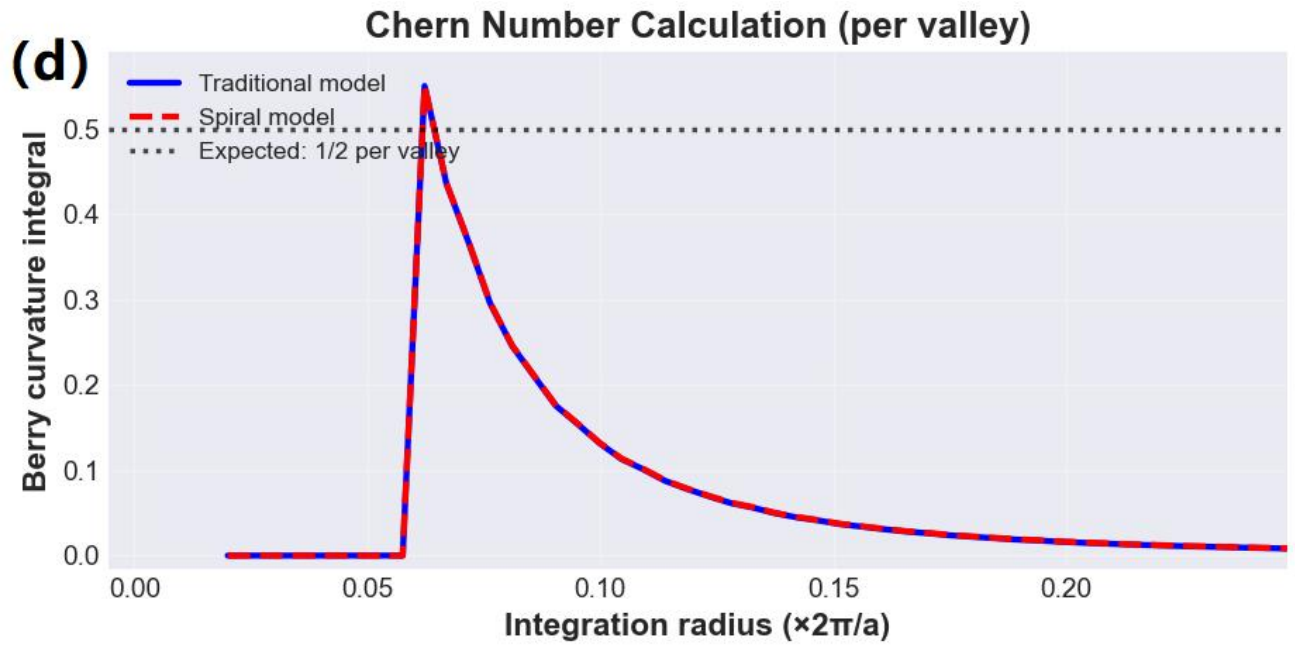

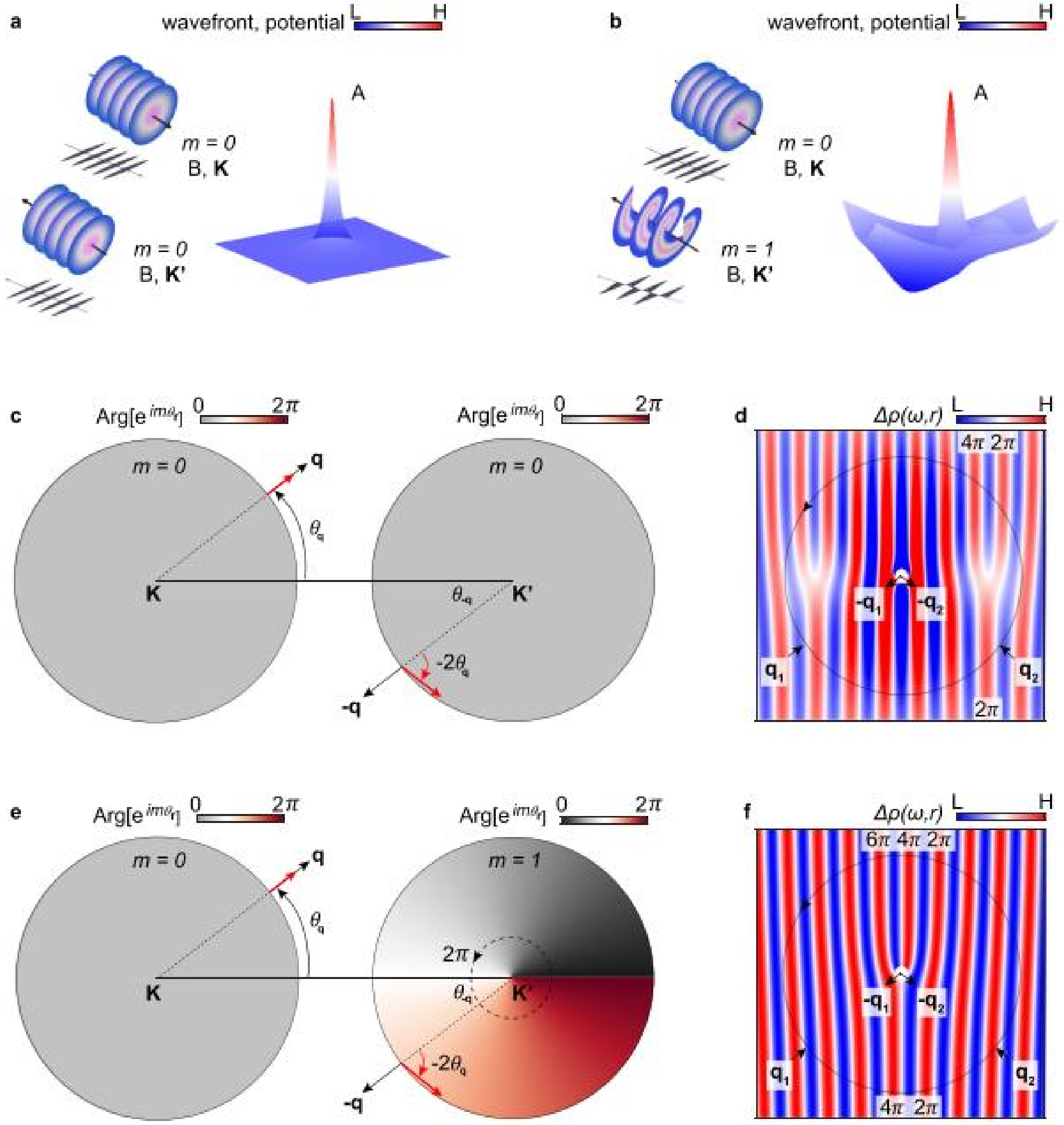

Fig. S2 Dynamics of intervalley scatterings and wavefront dislocations. **a** Aschematic diagram of intervalley interference between the incident wave from position **B** in **K** valley and the reflected wave back to position **B** in **K'** valley due to the defect potential A. The orbital angular momentum ***m*** is zero for these wavefronts. The color scale represents the intensity distributions for both wavefronts and the potential. **b** A wave scattered by a defect potential accompanied by a spread rotationally asymmetric potential, leading to reflection as a spiral wavefront with a finiteorbital angular momentum. **c** A intervalley scattering process for panel a between twonearest **K** and **K'** valleys, involving the pseudospin rotation. Here $\theta r$ is the polar angleof the position vector r. Black and red arrows denote the vector of momentum ***q*** andpseudospin, respectively. $\theta q$ is rotation angle of momentum. The color

gradient from black to red represents the variation of the phase. **d** A schematic diagram to show wavefront dislocations in total local density of states(LDOS) with the contribution of pseudospin rotation. Panels a and c are the partial processes of panel d. The q in all the directions scatters with the potential in the middle point and contributes to the accumulation of the phase for the closed loop. Two scattering directions $q_1$ and $q_2$ are shown in the figure. The phases from 2 π to 4π are marked in one of the wavefront dislocations and the value of the wave is donated by different colors. **e** A intervalley scattering process between two nearest **K** and **K'** valleys for panel b, involving both pseudospin and orbital angular momentum. The orbital angular momentum effect on valley manifests as the colored distribution with a varied phase. **f** The wavefront dislocation in LDOS modulation $\Delta\rho(\Delta K, r) \propto \cos(\Delta K \cdot r + \theta_r)$ with the contribution from both pseudospin rotation and orbital angular momentum coupling. The white dots in d and f schematically denote the defect position.Reference: Y.-W. Liu, Y.-C. Zhuang, Y.-N. Ren, C. Yan, X.-F. Zhou, Q. Yang, Q.-F. Sun, L. He, Visualizing a single wavefront dislocation induced by orbital angular momentum in graphene, Nature Communications, 15 (2024) 3546.

## 7. Comparison between BBC model and DFT calculation results

The electronic properties and structural relaxation of the graphene material were modeled using the Cambridge Sequential Total Energy Package(CASTEP). These calculations focused on the energetics and electronic properties of the graphene material. The Cut-off energy, band gaps, *k*-point grid used in this study are listed in Table S1. All the structures are fully optimized without any symmetry constraints until the force is smaller than 0.01 eV/Å and the energy tolerances less than $5.0 \times 10^{-6}$eV per atom. The convergence threshold of $1.0 \times 10^{-6}$eV per atom is selected in the self-consistent field calculations.

**Table S1** Cut-off energiesand *k*-points of graphene.

| | Cut-off energy | *k*-point |
|---|---|---|
| LDA | 750 eV | 7×7×1 |
| PBE | 440 eV | 7×7×1 |

| HSE06 | 750 eV | 7×7×1 |
|---|---|---|

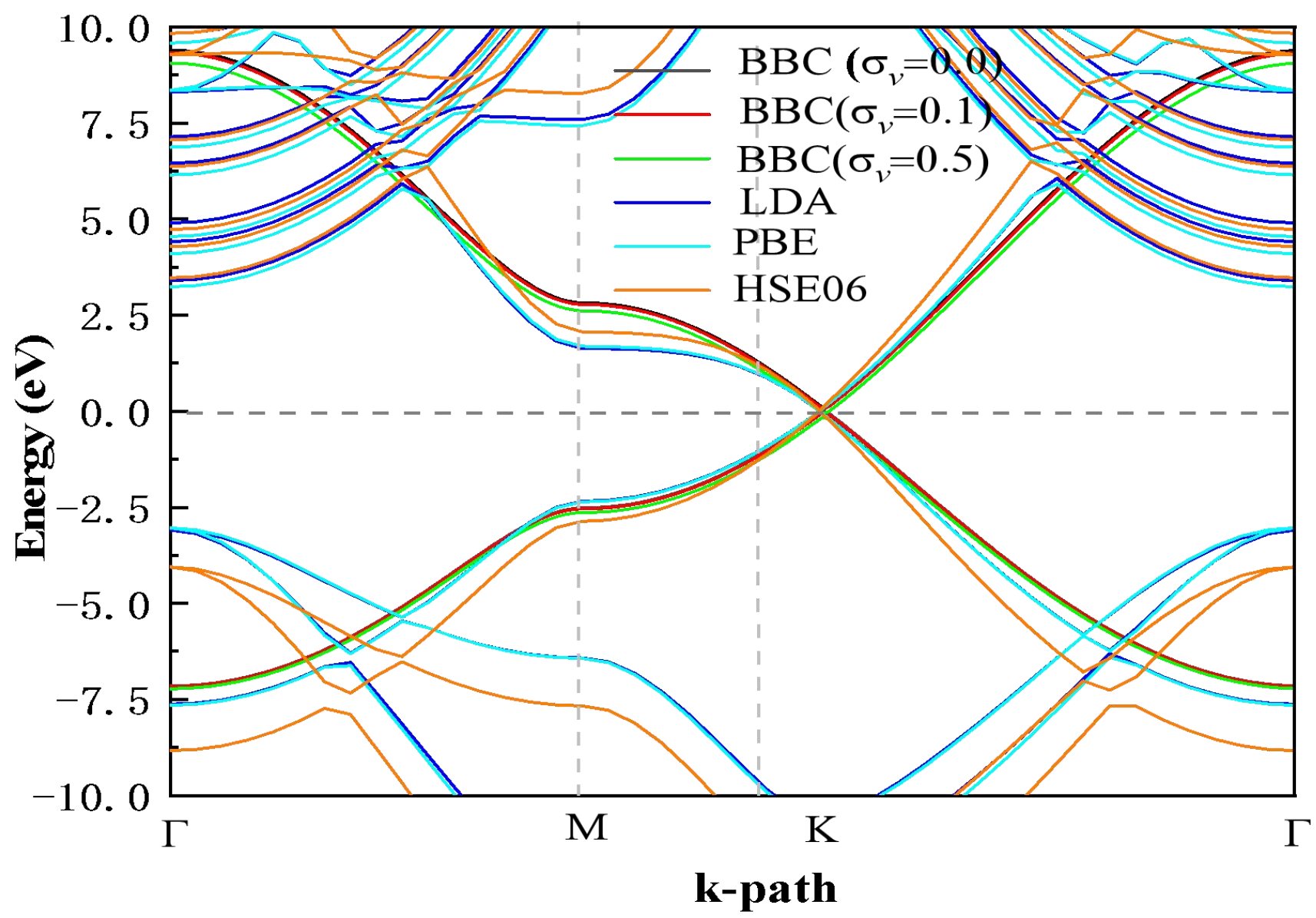

Fig. S3 Comparison of Band Structures between the BBC Model and DFT Calculations